\documentclass[twoside,twocolumn,9pt]{article}
\usepackage{extsizes}
\usepackage[super,sort&compress,comma]{natbib} 
\usepackage[version=3]{mhchem}
\usepackage[left=1.5cm, right=1.5cm, top=1.785cm, bottom=2.0cm]{geometry}
\usepackage{balance}
\usepackage{mathptmx}
\usepackage{amssymb}
\usepackage{sectsty}
\usepackage{graphicx} 
\usepackage{lastpage}
\usepackage[format=plain,justification=justified,singlelinecheck=false,font={stretch=1.125,small,sf},labelfont=bf,labelsep=space]{caption}
\usepackage{float}
\usepackage{fancyhdr}
\usepackage{fnpos}
\usepackage[english]{babel}
\addto{\captionsenglish}{%
  
}
\usepackage{array}
\usepackage{droidsans}
\usepackage{charter}
\usepackage[T1]{fontenc}
\usepackage[usenames,dvipsnames]{xcolor}
\usepackage{setspace}
\usepackage[compact]{titlesec}
\usepackage{hyperref}

\usepackage{epstopdf}

\definecolor{cream}{RGB}{222,217,201}

\newcommand{\add}[1]{\textcolor{black}{ #1 }} 
\newcommand{\MDGrevise}[1]{\textcolor{black}{ #1 }} 

\newcommand{\Ca}{\mathrm{Ca}}

\newcommand{\kb}{\hat{K}_B}

\usepackage{stfloats}
\usepackage{amsmath}

\begin{document}

\pagestyle{fancy}
\thispagestyle{plain}
\fancypagestyle{plain}{
\renewcommand{\headrulewidth}{0pt}
}

\makeFNbottom
\makeatletter
\renewcommand\LARGE{\@setfontsize\LARGE{15pt}{17}}
\renewcommand\Large{\@setfontsize\Large{12pt}{14}}
\renewcommand\large{\@setfontsize\large{10pt}{12}}
\renewcommand\footnotesize{\@setfontsize\footnotesize{7pt}{10}}
\makeatother

\renewcommand{\thefootnote}{\fnsymbol{footnote}}
\renewcommand\footnoterule{\vspace*{1pt}%
\color{cream}\hrule width 3.5in height 0.4pt \color{black}\vspace*{5pt}} 
\setcounter{secnumdepth}{5}

\makeatletter 
\renewcommand\@biblabel[1]{#1}            
\renewcommand\@makefntext[1]%
{\noindent\makebox[0pt][r]{\@thefnmark\,}#1}
\makeatother 
\renewcommand{\figurename}{\small{Fig.}~}
\sectionfont{\sffamily\Large}
\subsectionfont{\normalsize}
\subsubsectionfont{\bf}
\setstretch{1.125} 
\setlength{\skip\footins}{0.8cm}
\setlength{\footnotesep}{0.25cm}
\setlength{\jot}{10pt}
\titlespacing*{\section}{0pt}{4pt}{4pt}
\titlespacing*{\subsection}{0pt}{15pt}{1pt}

\fancyfoot{}
\fancyfoot[LO,RE]{\vspace{-7.1pt}\includegraphics[height=9pt]{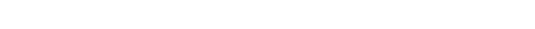}}
\fancyfoot[CO]{\vspace{-7.1pt}\hspace{13.2cm}\includegraphics{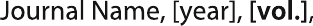}}
\fancyfoot[CE]{\vspace{-7.2pt}\hspace{-14.2cm}\includegraphics{head_foot/RF}}
\fancyfoot[RO]{\footnotesize{\sffamily{1--\pageref{LastPage} ~\textbar  \hspace{2pt}\thepage}}}
\fancyfoot[LE]{\footnotesize{\sffamily{\thepage~\textbar\hspace{3.45cm} 1--\pageref{LastPage}}}}
\fancyhead{}
\renewcommand{\headrulewidth}{0pt} 
\renewcommand{\footrulewidth}{0pt}
\setlength{\arrayrulewidth}{1pt}
\setlength{\columnsep}{6.5mm}
\setlength\bibsep{1pt}

\makeatletter 
\newlength{\figrulesep} 
\setlength{\figrulesep}{0.5\textfloatsep} 

\newcommand{\topfigrule}{\vspace*{-1pt}%
\noindent{\color{cream}\rule[-\figrulesep]{\columnwidth}{1.5pt}} }

\newcommand{\botfigrule}{\vspace*{-2pt}%
\noindent{\color{cream}\rule[\figrulesep]{\columnwidth}{1.5pt}} }

\newcommand{\dblfigrule}{\vspace*{-1pt}%
\noindent{\color{cream}\rule[-\figrulesep]{\textwidth}{1.5pt}} }

\makeatother

\twocolumn[
  \begin{@twocolumnfalse}
{\includegraphics[height=30pt]{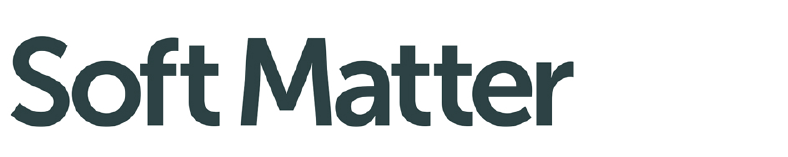}\hfill\raisebox{0pt}[0pt][0pt]{\includegraphics[height=55pt]{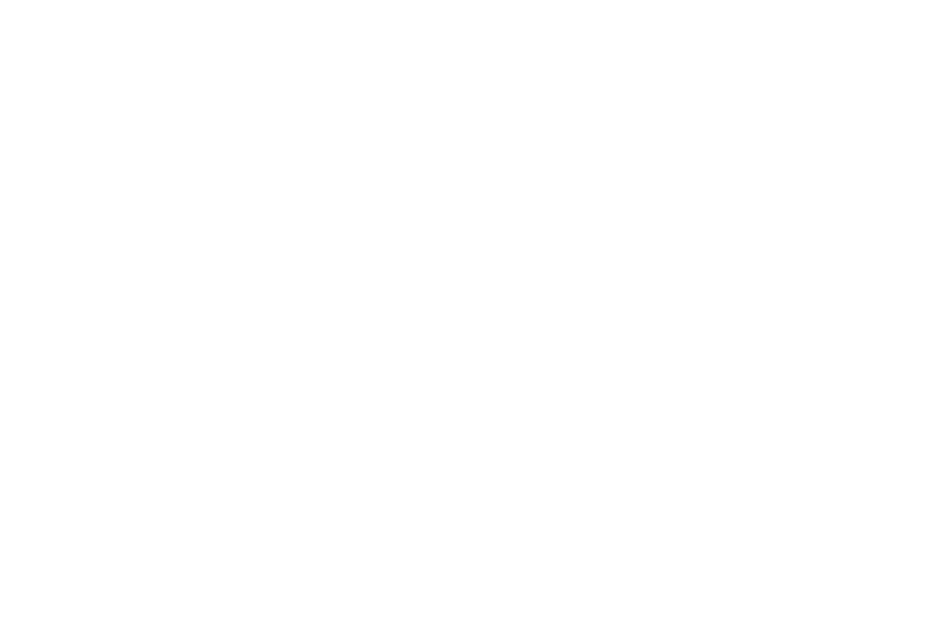}}\\[1ex]
\includegraphics[width=18.5cm]{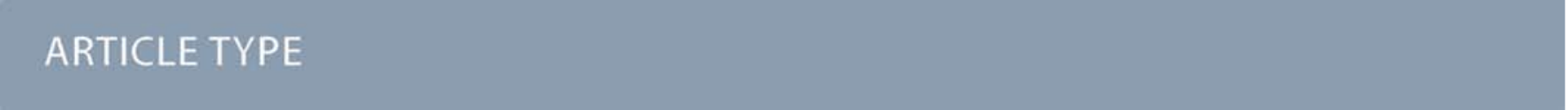}}\par
\vspace{1em}
\sffamily
\begin{tabular}{m{4.5cm} p{13.5cm} }

\includegraphics{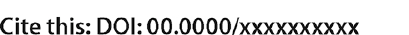} & \noindent\LARGE{\textbf{Coil-stretch-like transition of elastic sheets in extensional flows}} \\
\vspace{0.3cm} & \vspace{0.3cm} \\

 & \noindent\large{Yijiang Yu$^{a}$ and Michael D. Graham$^{a\ast}$} \\

\includegraphics{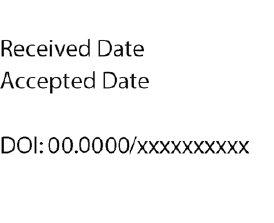} & \noindent\normalsize{The conformation of a long linear polymer dissolved in fluid and exposed to an extensional flow is well-known to exhibit a ``coil-stretch" transition, which for sufficiently long chains can lead to bistability. The present work reports computations indicating that an analogous ``compact-stretched" transition arises in the dynamics of a thin elastic sheet.  Sheets of nominally circular, square or rectangular shape are simulated in planar and biaxial flows using a finite element method for the sheet conformations and a regularized Stokeslet method for the fluid flow.  If a neo-Hookean constitutive model is used for the sheet elasticity, the sheets will stretch without bound once a critical extension rate, as characterized nondimensionally by a capillary number, is exceeded. Nonlinear elasticity, represented with the Yeoh model, arrests the stretching, leading to a highly-stretched steady state once the critical capillary number is exceeded. For all shapes and in both planar and biaxial extension, a parameter regime exists in which both weakly stretched (compact) and strongly stretched states can be found, depending on initial conditions. I.e. this parameter regime displays bistability. 
\MDGrevise{As in the long-chain polymer case, the bistable behavior arises from the hydrodynamic interaction between distant elements of the sheet, and vanishes if these interactions are artificially screened by use of a Brinkman model for the fluid motion.}
 While the sheets can transiently display wrinkled shapes, all final shapes in planar and biaxial extension are planar.} \\
\end{tabular}

 \end{@twocolumnfalse} \vspace{0.6cm}

  ]

\renewcommand*\rmdefault{bch}\normalfont\upshape
\rmfamily
\section*{}
\vspace{-1cm}


\footnotetext{\textit{$^{a}$~Department of Chemical \& Biological Engineering, University of Wisconsin-Madison,1415 Engineering Drive, Madison, WI 53706, USA. }}
\footnotetext{\textit{$^{*}$~mdgraham@wisc.edu}}
%
%



\section{Introduction}
In recent years, thin-structured materials have gained considerable attention in many fields. For example, freely suspended sheets immersed in a fluid environment arise the synthesis and processing of polymer networks \cite{lee2013, ni2015}. Polymer sheets are often used as planar substrates in soft origami-inspired devices, which further assemble into complicated structures\cite{so2017,jamal2011, hubbard2017,na2015}. Thin polymer films are also used as the basis of self-propelled swimming devices, with living cells deposited on the sheet to drive the motion\cite{feinberg2007}. \add{Highly extensible polymer films, hydrogels structures and their composites have shown great potential in biomedical and electronic applications\cite{chun2013free, zhao2019highly, gaharwar2011highly, liu2012synthesis, mehta2018engineering}.} Many of above examples involve complicated fluid-structure interactions, but little is known about the dynamics of soft sheet-like particles in fluid.	The present work aims to shed light on some aspects of these dynamics through simulations of soft elastic sheets in two canonical flow fields that arise in many applications such as coating operations: planar and biaxial extensional flows.


While soft elastic sheets have not been studied in these flows, long-chain polymer molecules have been.  
An important phenomenon in this situation is the coil-stretch transition\cite{de1974,schroeder2003,fuller1981}: polymers show abrupt change in conformation from a coil state to a strongly stretched state when reaching a critical flow strength in planar extension.
The discontinuity in the stretched length marks a bistable regime that corresponds to the existence of multiple stable (or more precisely, metastable) conformations. This behavior was first predicted by De Gennes in 1974\cite{de1974}, based on the idea that, compared to a fully stretched state,  in the coiled state the inner monomers are hydrodynamically screened and therefore less exposed to the strong stretching flow. Fuller and Leal modeled this effect using a bead-spring dumbbell  with conformation-dependent drag on the beads, showing that this simple approach could yield coil-stretch hysteresis \cite{fuller1981}. Schroeder \textit{et al.} experimentally confirmed this prediction, visualizing the hysteresis behavior with fluorescent dyed Escherichia coli DNA molecules in dilute solution under planar extension\cite{schroeder2003}. They also probed the hysteresis phenomenon with a Brownian dynamics simulation of a bead-spring polymer model and found a double-welled effective potential of polymer conformational energy around the transition flow strength that leads to hysteresis in conformation. The coil-stretch transition and hysteresis are also observed in simulations beyond the dilute limit. Mohammad \textit{et al.} applied molecular dynamics simulation to study atomistic entangled polyethylene melts in planar extension\cite{nafar2018}. They found that the bistable region in the polymer melts is wider than in  dilute solution, with a higher transition flow strength.

Similar conformational transitions have been observed in systems other than polymers as well. 
Kantsler \textit{et al.} performed experiments to study the stretching of a single tubular vesicle suspended in planar extensional flow\cite{kantsler2008}. They found a transition in conformation from a tubular to a dumbbell shape when reaching a critical extension rate. With further increasing flow strength, the shape evolves to a pearling state due to unstable higher order shape modes. This behavior is also found in a computational study by Narsimhan \textit{et al.}\cite{narsimhan2015}, who also considered other extensional flows. Kumar \textit{et al.} recently developed a microfluidic Stokes trap to make a vesicle remain at the stagnation point and extended the work of Kantsler \textit{et al.} by giving a more detailed analysis of the vesicle shape transition in the parameter space of reduced volume and bending modulus\cite{kumar2020}.


 In contrast to the substantial literature on polymers or vesicles, there is only a small amount of work on the dynamics of deformable freely suspended sheets in flow \MDGrevise{and no prior study has probed the possibility of hysteretic conformational transitions of extensible sheets in extensional flow}. (Though there exist many studies on flapping sheets with a clamped edge in flow (flapping flags) \cite{shelley2011flapping, alben2008flapping}, this is not our interest here.)  
Motivated by applications of relatively inextensible sheetlike materials such as graphene and boron nitride, Green and Xu investigated the dynamics of freely suspended nanosheets in both extensional and shear flow \cite{xu2014, xu2015}. They modeled a nominally square sheet with a bead-rod network inspired by coarse-grained models of polymers in solution, simulating the model with a Brownian dynamics method. Because of the inextensibility of the rods, the model they consider is highly resistant to stretching deformations, so much of the dynamics in this system are determined by the relative importance of flow, bending resistance and Brownian motion. In biaxial extensional flow, a sheet with large bending stiffness tends to stay at a flat conformation. With small bending stiffness, Brownian motion leads to crumpling \cite{xu2015}. In shear flow at low bending stiffness, they found that sheets undergo a cyclic crumple-stretch-crumple motion \cite{xu2014}. 	
 Botto and co-workers performed molecular dynamic simulations to study the liquid phase exfoliation of graphite sheets in shear. They studied the critical shear rate for exfoliation to occur in different solvent and developed a theoretical model that can predict the critical shear rate in simulation\cite{gravelle2020liquid}. They also applied molecular dynamics simulations to explore the effects of hydrodynamic slip on the dynamics of a sheet in shear \cite{kamal2020hydrodynamic}. Interestingly, in the presence of slip, they found that a rigid nanosheet can take on a stable steady orientation with a finite orientation angle with respect to the flow direction. Finally, Dutta and Graham investigated the dynamics of a piecewise rigid creased sheet -- a model of an origami figure known as a Miura pattern \cite{dutta2017}. Depending on geometry and initial conditions, this sheet can show steady, periodic or quasiperiodic orientational and conformational dynamics in shear.

The present work describes the behavior of sheetlike particles in a parameter regime that has not been explored in the studies above. We consider soft (highly extensible) elastic sheets, taking Brownian forces to be negligible compared to hydrodynamic and elastic ones, and focus on planar and biaxial extensional flows. 
We consider three different rest shapes: disc, square, and rectangle. 
The rest of the paper is organized as follows: Section \ref{sec:methods} summarizes the model we use and the discretization and solution methods. 
Section \ref{sec:results}  shows detailed results for the sheet dynamics, illustrating a bistability phenomenon analogous to the coil-stretch transition. Concluding remarks are presented in Section \ref{sec:conclusion}.  

\section{Model and Methods}\label{sec:methods}
\subsection{Model formulation}
We consider a very thin, neutrally buoyant, elastic sheet suspended in an unbounded , incompressible Newtonian fluid with viscosity $\eta$ and density $\rho$, and subjected to planar and biaxial extensional flow fields.  Expressed in Cartesian coordinates, the fluid velocity $\mathbf{v}_\infty$ in the absence of the particle is given in planar extension by $\mathbf{v}_\infty = \dot\epsilon\left[x, 0, -z \right]^T$ and in biaxial extension by $\mathbf{v}_\infty = \dot\epsilon\left[x/2, y/2, -z \right]^T$, with $\dot{\epsilon}$ the strain rate. 

\MDGrevise{The sheet is modeled as an elastic continuum. Fluid cannot pass through the sheet, and each point on the sheet is taken to move with the local velocity of the fluid. That is, the usual no-penetration and no-slip boundary conditions are applied for the fluid velocity at the sheet surfaces. In this case, no specification of the details of fluid-solid interfacial forces is necessary. As noted above, for atomically smooth surfaces, the no-slip boundary condition may not apply, in which case interesting dynamics have been seen to arise in simulations of rigid sheets in shear flow\cite{kamal2020hydrodynamic}. That case is not considered here. }
 In this work, we focus on sheets with three rest shapes:  (a) disc with radius $a$, (b)  square with edge $2a$ and (c)  rectangle with short edge $a$ and long edge $2a$. The rest thickness $h$ is much less than $a$. 
 The sheets are taken to be sufficiently thin that the fluid exerts no traction along their edges. 

The mechanical response of a very thin sheet is split into two parts: in-plane shearing elasticity, with strain energy $E_s$, and out-of-plane bending elasticity, with strain energy $E_b$. The total strain energy is thus
\begin{equation}
    E = E_s + E_b.
\end{equation}
The in-plane shearing energy $E_s$ on the sheet surface $\Gamma$ can be written as
\begin{equation}
    E_s = \int_\Gamma W\;dS,
\end{equation} 
where $W$ is the areal shear strain energy density \add{that is defined pointwise on the sheet surface}. 
We consider the sheet as isotropic and incompressible, and describe the in-plane energy with the Yeoh model, a model used to simulate rubber-like materials\cite{yeoh1993}. Its strain elastic energy $W$ is given by:
\begin{equation}
    W = \frac{G}{2}\left( \lambda_1^2 + \lambda_2^2 + \lambda_3^2 -3\right) + cG\left( \lambda_1^2 + \lambda_2^2 + \lambda_3^2 -3\right)^3.
    \label{eq:yeoh}
\end{equation}

Here, $G$ is a two-dimensional shear elasticity modulus given by $G_{\mathrm{3D}}h$, where $G_{\mathrm{3D}}$ is the usual $\mathrm{3D}$ shear modulus. The quantities $\lambda_1$ and $\lambda_2$ are the \MDGrevise{local} principal stretch ratios along the \MDGrevise{(local)} tangential directions of the sheet surface, while the normal stretch ratio $\lambda_3$ is determined by incompressibility of the sheet material: $\lambda_1\lambda_2\lambda_3 = 1$. The parameter $c$ determines the weight of the cubic terms in the energy formulation and determines whether the material is strain-softening or strain-hardening. When $c = 0$, the Yeoh model reduces to the strain-softening neo-Hookean (NH) model. 

For bending energy, we apply a simple linear bending model that penalizes deflection between neighboring discretized elements.\cite{Fedosov2010,fedosov2010multiscale} Details of the discretization will be covered in later paragraphs. This energy has the form
\begin{equation}
    E_b = \sum_{adjacent\ \alpha,\beta}k_b \left[1-\cos(\theta_{\alpha\beta} - \theta_0)\right],
\end{equation}
where $k_b$ is a bending constant, $\theta_{\alpha\beta}$ is the angle between two neighboring elements and $\theta_{0}$ is the angle at equilibrium. In this study, we assume the sheet is flat at rest, so $\theta_{0} = 0$ for all adjacent element pairs. The bending constant $k_b = \sqrt{3}(2K_B+\overline{K_B})$ denotes an averaged bending modulus, which is  derived from the Helfrich bending energy of a spherical shell with a zero spontaneous curvature\cite{Fedosov2010}. Here $K_B$ is the bending modulus and is related to the Gaussian curvature modulus $\overline{K_B}$ by  $\overline{K_B}/K_B = \nu -1$, where the Poisson ratio $\nu = 1/2$ for an incompressible sheet\cite{timoshenko1959theory}. A nondimensional bending stiffness can be defined: $\hat K_B = K_B/a^2G$,

The simulations performed and the results presented are in nondimensional form. Lengths are scaled with $a$ and time with the inverse extension rate $1/\dot\epsilon$. Forces are nondimensionalized by viscous drag force $\eta\dot\epsilon a^2$. With these choices, two nondimensional parameters arise that characterize the deformability of the sheet by the flow: the in-plane deformability is determined by the capillary number $\Ca = \eta\dot\epsilon a/G$, \MDGrevise{with the competition between flow and bending deformations characterized by $\Ca/\hat{K}_B$}. It is worth pointing out here that in all cases studied here, the sheet becomes flat at long times, in which case the bending energy becomes irrelevant. It does, however, have an effect on dynamics at short times, determining the transient wrinkling behavior that the sheet exhibits while orienting with the flow.


\add{Our results below indicate that complex behavior in extensional flow occurs once $\Ca \gtrsim 0.1$. We present here some estimates to illustrate how these results will be relevant to microscale sheets of specific materials.   Materials like  polymer hydrogels usually have shear modulus $G_{3D} \approx 1 \mathrm{kPa}$) \cite{gaharwar2011highly, liu2012synthesis, mehta2018engineering}. If a disc-shaped polyethylene glycol (PEG) hydrogel sheet with radius of 100 $\mathrm{\mu m}$ and thickness of 100 nm is suspended in water  ($\eta \sim 10^{-3} \;\mathrm{Pa\cdot s}$), a modest extension rate of about $10^2\mathrm{s^{-1}}$ will yield $\Ca = 0.1$.} 

\MDGrevise{Other interesting two-dimensional materials, such as graphene flakes, have much too large a shear modulus ($G_{3D} \approx 300\; \mathrm{GPa}$\cite{liu2012shear}) to yield a capillary number in the range where they will be strongly stretched by flow, except in extreme conditions. }

\add{In addition, the thermal fluctuation for polymer sheets at room temperature is considered negligible compared to strain energy. If we still consider x`the above PEG hydrogel, its estimated shear strain energy ($\sim 10^9 k_BT$) and bending energy ($\sim 3\times10^5 k_BT$) are much larger than thermal energy. Therefore, thermal fluctuations will negligible influence on the dynamics.}  
    
\subsection{\MDGrevise{Numerical description of sheet elasticity}}

The numerical method for the elasticity problem is adapted from Charrier $\textit{et al.}$ \cite{Charrier1989,Pappu:2008in}. \add{We simulate the sheet by keeping track of material points as nodes on the sheet surface.} The sheet surface is discretized into triangular elements with a node at each corner. 
From the above two energies, we obtain the elastic force $\mathbf{F}_e$ exerted on each node \add{from} the first variation of the total energy with respect to the nodal displacements. 
Each discretized element of the sheet is assumed to have homogenous deformation, so the element edges always remain linear. The deformed element is compared to its equilibrium shape under a local coordinate transformation via a rigid body rotation and the displacement for any point inside the element is obtained by linear interpolation from nodes. 
The detailed implementation can be found in references\cite{Kumar:2012ev,Fedosov2010,fedosov2010multiscale}. 
\add{The total nodal force $\mathbf{F}_e$ is evaluated by summing the elastic force $\mathbf{F}_{e,i}$ due to deformation of each surrounding element shared by the node: $\mathbf{F}_e = \sum_i \mathbf{F}_{e,i}$, where the sum is over all elements meeting at the node.}
For the results shown, we discretize the disc with 1600 elements and 841 nodes, the square with 2048 elements and 1089 nodes, and the rectangle with 1024 elements and 561 nodes. We have verified that changes in mesh resolution lead to only small quantitative changes in the results and no qualitative changes.

\subsection{\MDGrevise{Numerical description of fluid motion}}
We consider the case of very small sheets, such that the particle Reynolds number $Re=\rho\dot\epsilon a^2/\eta$ is much less than unity and the fluid motion is governed by the Stokes equation. Neglecting the inertia of the sheet, the total elastic force $\mathbf{F}_e$ exerted on each node by the surrounding solid is balanced by the force $\mathbf{F}_h$ exerted by the fluid on the node:
\begin{equation}
    \mathbf{F}_e + \mathbf{F}_h =\mathbf{0}.
\end{equation}
Thus, the sheet exerts a force on the fluid at each nodal position. 
To account for the fact that the forces are not completely localized to the nodal positions, we use the method of regularized Stokeslets \cite{Cortez2005}: the force $\mathbf{F}_i$ exerted by node $i$ on the fluid corresponds to a regularized force density $\mathbf{f}^\kappa = \mathbf{F} \delta_\kappa(\mathbf{x}-\mathbf{x}_i)$, where $\delta_\kappa(\mathbf{x})$ is a regularized delta function with regularization parameter $\kappa$. Thus, the governing equations are the Stokes equation with regularized nodal forces and the continuity equation:

\begin{equation}
    \begin{array}{c}
        {-\nabla p+\eta \nabla^{2} \mathbf{v}+\sum_i\mathbf{F}_i \delta_\kappa(\mathbf{x})=\mathbf{0}} \\ 
        {\nabla \cdot \mathbf{v}=0}.
    \end{array}
\end{equation}

Here the sum is over the nodal positions.
The velocity field generated due to a regularized point force $\mathbf{f}= \mathbf{F}\delta_\kappa(\mathbf{x})$ can be represented using a regularized Stokeslet $\mathbf{G}_\kappa$:
\begin{equation}
    \mathbf{v}_\kappa(\mathbf{x}) = \mathbf{G}_\kappa(\mathbf{x})\cdot \mathbf{F}.
\end{equation}
As $1/\kappa\rightarrow 0$, $\mathbf{G}_\kappa$ reduces to the usual Stokeslet operator $\mathbf{G}(\mathbf{x})=1/8\pi\eta r(\mathbf{I}+\mathbf{x}\mathbf{x}/r^2)$, where $r=|\mathbf{x}|$.
There are many ways to regularize a delta function $\delta_\kappa(\mathbf{x})$; we choose a regularization function for which the deviation between $\mathbf{G}$ and $\mathbf{G}_\kappa$ decays exponentially as $\kappa r\rightarrow\infty$ \cite{hernandez2007,Graham2018}:

\begin{equation}
    \delta_\kappa(\mathbf{x}) = \frac{\kappa^3}{\sqrt{\pi}^3}\exp(-\kappa^2r^2)\left[\frac52-\kappa^2r^2\right].
\end{equation}
With this choice, 
\begin{equation}
    \mathbf{G}_{\kappa}(\mathbf{x})=\frac{\operatorname{erf}(\kappa r)}{8 \pi \eta r}\left(\mathbf{I}+\frac{\mathbf{xx}}{r^{2}}\right)+\frac{\kappa e^{-\kappa^{2} r^{2}}}{4 \pi^{3 / 2} \eta}\left(\mathbf{I}-\frac{\mathbf{xx}}{r^{2}}\right).
\end{equation}

In the simulations, $\kappa$ must be chosen to scale with the minimum node-to-node distance $l_{\mathrm{min}}$. We take $\kappa l_{\mathrm{min}} = 2.1842$, which is obtained using a validation case discussed below. 
\add{Due to the no-slip and the no-penetration boundary conditions, the velocity of each node on the sheet surface equals the fluid velocity at the nodal position.}
\MDGrevise{This velocity is the sum} of the free velocity $\mathbf{v}_{\infty}$ in the absence of the sheet and the perturbation velocity $\mathbf{v}_{p}$ generated by the nodal forces:
\begin{equation}
    \mathbf{v}(\mathbf{x}) =\mathbf{v}_\infty(\mathbf{x}) +\mathbf{v}_p(\mathbf{x})= \mathbf{v}_\infty(\mathbf{x}) +\sum_i\mathbf{G}_\kappa(\mathbf{x}-\mathbf{x}_i)\cdot \mathbf{F}_i.
    \label{eq:v}
\end{equation}
\MDGrevise{The position of each node is determined by integrating the velocity with a fourth-order Runge-Kutta method.}
The time step applied in simulation follows $\Delta t = 0.1\Ca l_{\mathrm{min}}$, with an upper limit of $5 \times 10^{-4}$ strain unit.

\subsection{Model validation}
\begin{figure}[!htb]
    \centering 
    \captionsetup{justification=raggedright}
    \includegraphics[width=0.5\textwidth]{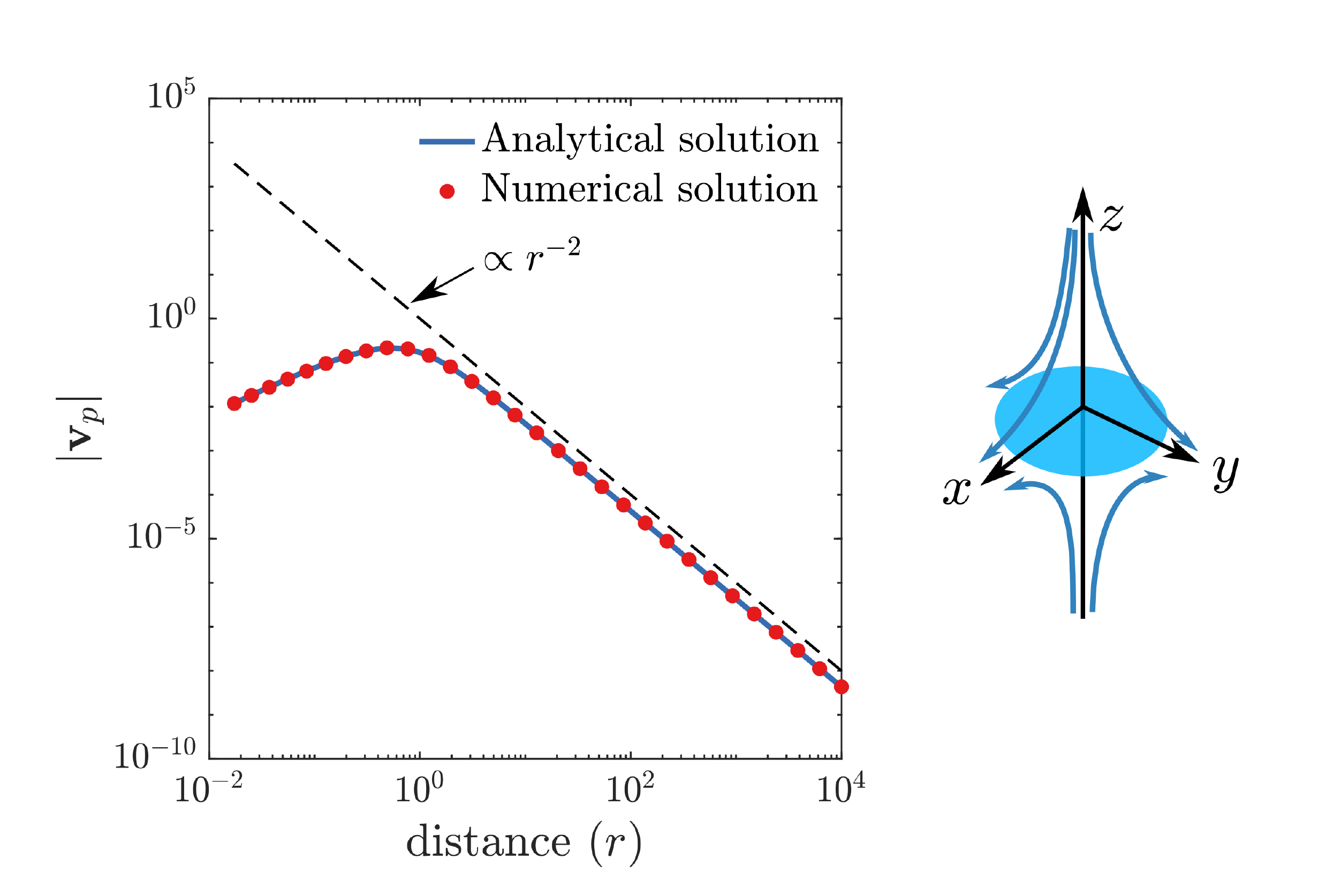}
    \caption{Perturbation velocity norm $|\mathbf{v}_p|$ vs. distance $r$ along a line away from the origin for a disc in biaxial flow. The line  chosen follows the vector $[1, 1, 100]$. The black dashed line is proportional to $r^{-2}$.}
    \label{fig:model_valid}
\end{figure}

Here we present a validation case for the regularized Stokeslet  formulation: a rigid disc with radius $m$ and no-slip boundary conditions lying in the stretching $x-y$ plane of unbounded biaxial extensional flow. We set $\Ca = 0.01$ for a disc (discretized by 1600 triangular elements with $l_{min} = 0.05$) to ensure negligible deformation on the sheet surface. For this case, an analytical solution can be found by solving the stream function form of the Stokes equation\cite{happel2012}. In cylindrical coordinates with $r$ and $z$ the usual radial and axial directions, the solution $\psi_p$ for the perturbation away from pure biaxial extension is
\begin{equation}
    \psi_p(r,z) = \frac{1}{\pi} r^2z\dot\epsilon\left(\cot^{-1}(\lambda)-\frac{\lambda}{\lambda^2+1}\right),
\end{equation}
 where
 \begin{equation}
 \lambda = \sqrt{\frac{r^2+z^2-m^2+\sqrt{(r^2+z^2-m^2)^2+4m^2z^2}}{2m^2}}.
 \end{equation}
Figure \ref{fig:model_valid} compares the perturbation velocity norm along an arbitrarily chosen path away from disc center, showing excellent agreement between the analytical and regularized Stokeslet solutions. The perturbation velocity decays as $r^{-2}$ because to leading order in $1/r$, a neutrally-buoyant particle  acts like a force dipole. 

\section{Results}\label{sec:results}

\subsection{Planar extensional flow}\label{sec:planar}

\begin{figure*}[b]
    \centering
    \captionsetup{justification=raggedright}
    \includegraphics[width=\textwidth]{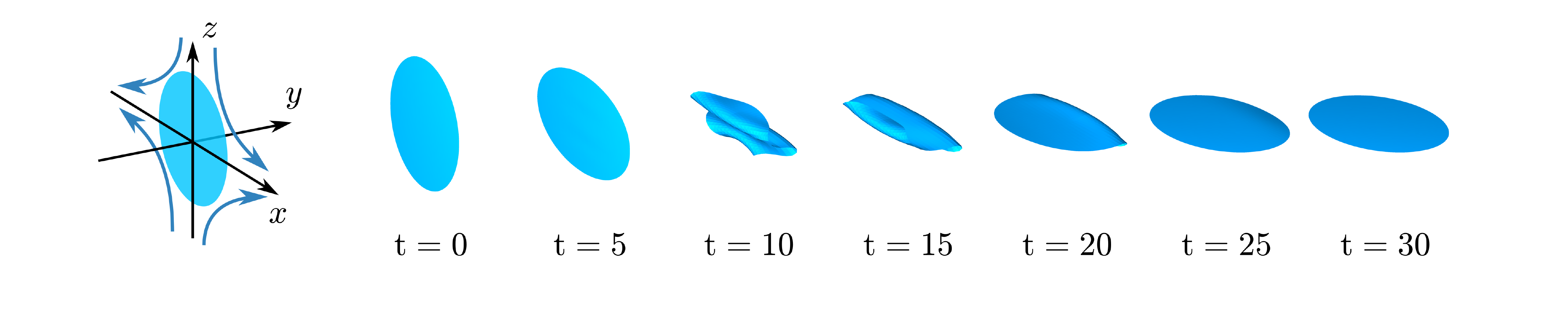}
    \caption{Time evolution of a deformable disc initially aligned with the $x-z$ plane in planar extensional flow ($\Ca = 0.2;\; \hat K_B = 1\times 10^{-3}$).}
    \label{fig:resp_disc_flip}
\end{figure*}

We begin the discussion of sheets in planar extensional flow with an important general observation. Consider an initial condition where the sheet is flat, with arbitrary orientation. We find that this initial condition will almost always evolve at long time to a flat shape aligned with the $x-y$ plane, where the $x$- and $y$-axes are the extensional and neutral directions for the flow, respectively. The only exception is the case where the sheet is initially oriented perfectly in the $x-z$ plane. By symmetry, this sheet's orientation will remain in this plane for all time. However, this situation seems to be unstable for a sheet of any shape with any finite deformability, as exemplified in Figure \ref{fig:resp_disc_flip} for a disc. In all cases studied, the transient dynamics, and in particular the degree of transient wrinkling as the sheet is compressed along the $z$-direction, depend on $\Ca$ and $\hat{K}_B$, but eventually a flat state aligned with the $x-y$ plane is reached. Accordingly, all subsequent results consider sheets with this orientation \add{ and with $\hat{K}_B=1\times 10^{-3}$. Only the in-plane deformability ($\Ca$) is varied in the rest of the paper, as $\kb$ has no influence on the final conformation.}

\begin{figure}[!htb]
    \centering
    \captionsetup{justification=raggedright}
    \includegraphics[width=0.5\textwidth]{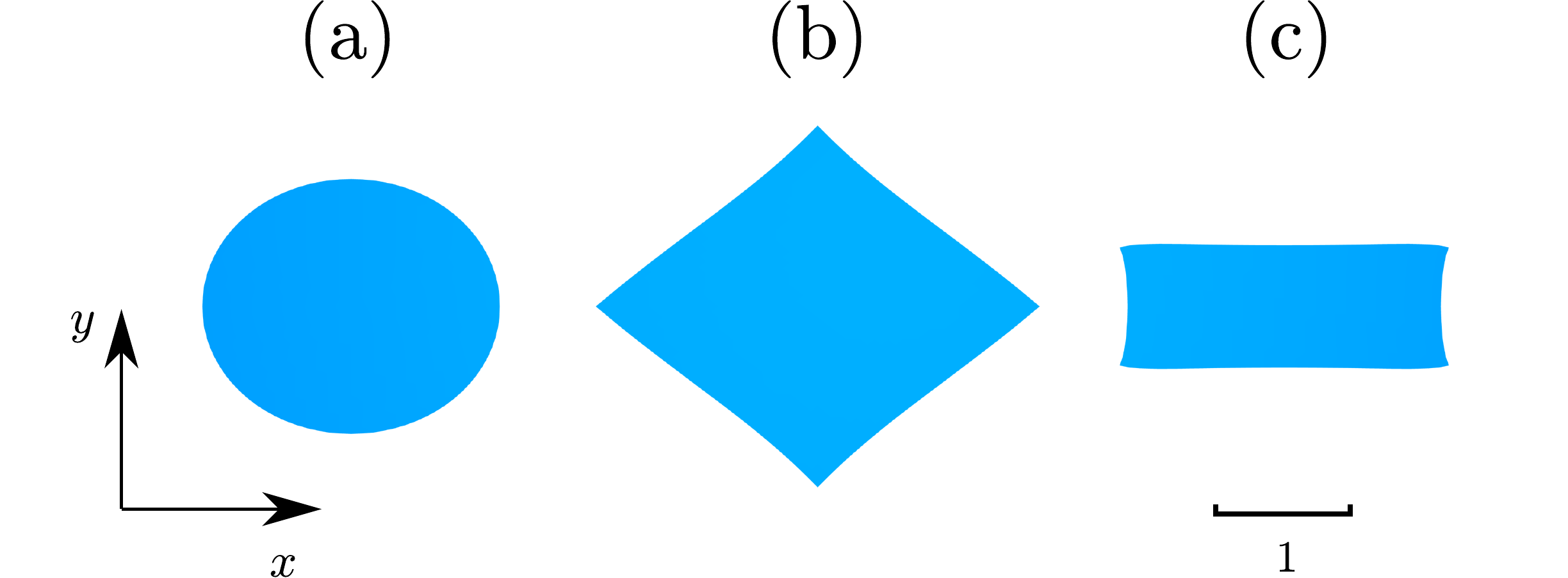}
    \caption{Stable steady-state conformation of a sheet in planar extensional flow with neo-Hookean elasticity at $\Ca = 0.1$: (a) disc. (b) a square. (c) rectangle. The scale bar 1 dimensionless length unit.}
    \label{fig:resp_steady_state_compact}
\end{figure}

Another stability issue arises for sheets whose initial shapes are not circular. Figure \ref{fig:resp_steady_state_compact} shows steady states for the three shapes we consider here, for low $\Ca$ and the neo-Hookean constitutive model (Eq.\ref{eq:yeoh} with $c=0$). For a disc, the steady state is roughly ellipsoidal, as shown in Figure \ref{fig:resp_steady_state_compact}a.  For a non-radially symmetric geometry like a square, we also need to consider its stable orientation. Even at very small  $\Ca$, an initial condition where the sides of the square are aligned with the $x$ and $y$ coordinate axes will eventually rotate until the diagonal aligns with the flow direction, reaching a steady state as indicated in Figure \ref{fig:resp_steady_state_compact}b. (At $\Ca=0$, where the sheet is perfectly rigid, Stokes flow reversibility would prevent this reorientation.) 
By contrast, the rectangular shape we study here remains symmetric at low $\Ca$, as illustrated in Figure \ref{fig:resp_steady_state_compact}c). 

\begin{figure}[!htb]
    \centering
    \captionsetup{justification=raggedright}
    \includegraphics[width=0.42\textwidth]{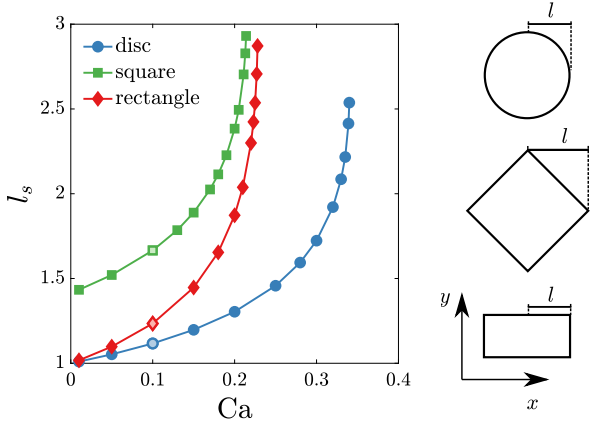}
    \caption{Steady-state stretched length $l_s$ vs. $\Ca$ for  deformable sheets with neo-Hookean elasticity in planar extensional flow. The data points filled with light color correspond to the shapes shown in Figure \ref{fig:resp_steady_state_compact}. The critical $\Ca$ for each geometry are: disc ($\Ca = 0.34$), square ($\Ca = 0.214$), rectangle ($\Ca = 0.228$). The subplot indicates how length $l$ is measured and the preferred stable orientation. }
    \label{fig:resp_all_singular}
\end{figure}

Continuing with the neo-Hookean model, Figure \ref{fig:resp_all_singular} shows the steady-state extension $l_s$ as a function of $\Ca$ for the three shapes.  In each case, there is a critical value of $\Ca$ beyond which the sheet will stretch without bound, not reaching a steady state shape. The appearance of the singularity is caused by the strain-softening property of the neo-Hookean model -- the increasing flow strength overcomes the elastic response of the sheet -- and is analogous to the behavior found in a bead-spring dumbbell model of a polymer molecule when the spring force obeys Hooke's law \cite{Graham2018} and the Weissenberg number exceeds a critical value. In the case of the dumbbell model,  if the Hookean spring model is replaced by a finitely extensible spring, the singularity vanishes and the chain reaches a steady state where the chain is highly stretched. This is the simplest version of the ``coil-stretch'' transition of flexible polymers. In the sheet case, we will refer to conformations at $\Ca$ values below the transition as ``compact". 

 \begin{figure}[!htb]
    \centering
    \captionsetup{justification=raggedright}
    \includegraphics[width=0.47\textwidth]{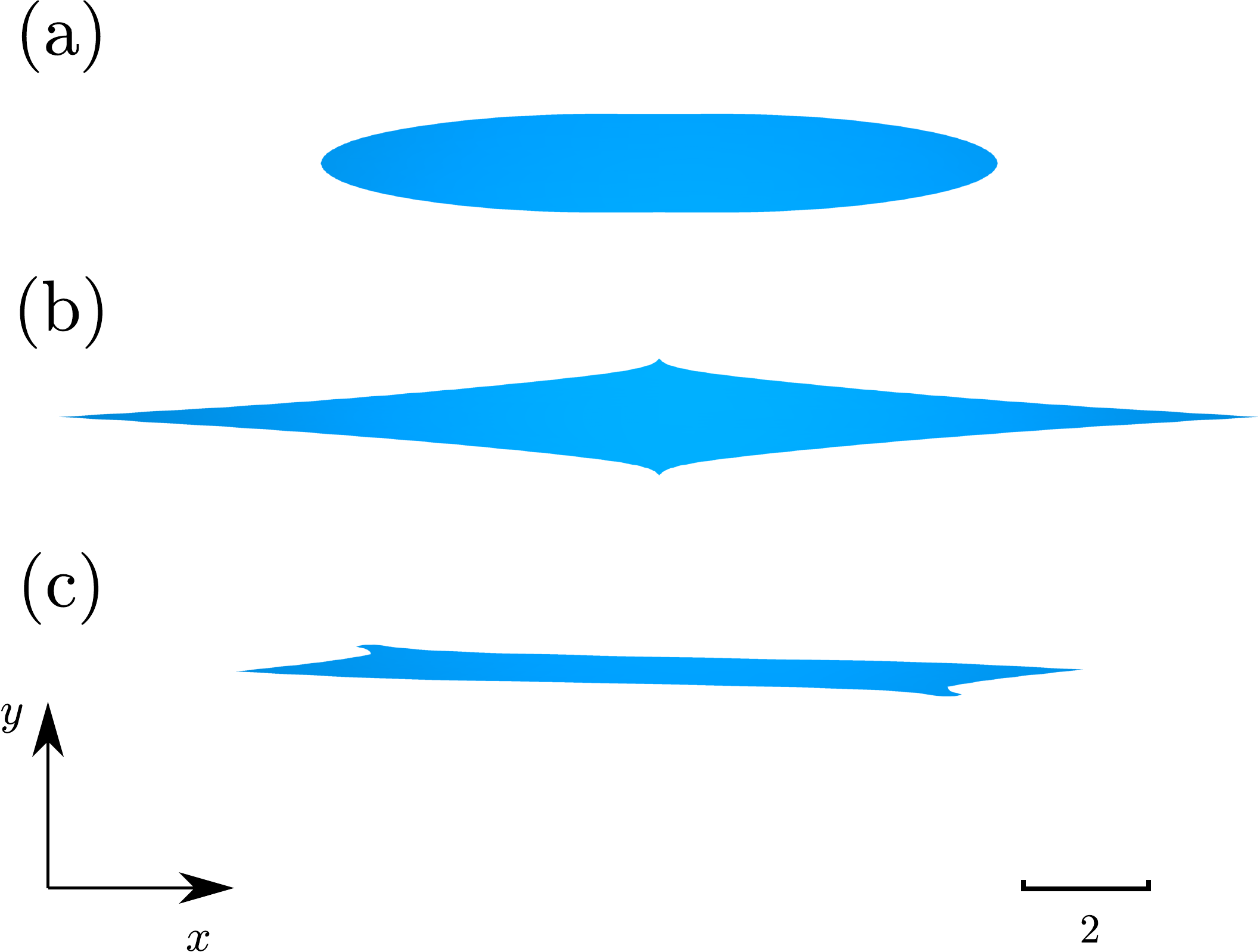}
    \caption{The steady-state conformation of a stretched sheet in planar extensional flow with Yeoh elasticity with $c = 5\times 10^{-5}$: (a) a disc at $\Ca = 0.35$. (b) a square at $\Ca = 0.22$. (c) a rectangle at $\Ca = 0.24$. The  scale bar is 2 dimensionless length units.}
    \label{fig:resp_steady_state_stretched}
\end{figure}

To further explore the compact-stretched transition, we turn to the strain-hardening Yeoh model.  At low $\Ca$, a finite, small value of $c$ leads to negligible change in conformation. However, once the critical capillary number is exceeded, the the behavior is very different: the singularity is replaced by a sudden jump in the stretched length within a small $\Ca$ interval around the critical $\Ca$.
Figure \ref{fig:resp_steady_state_stretched} shows steady states for our three rest shapes with $c=5\times 10^{-5}$ at values of $\Ca$ just above the singularity in the neo-Hookean case, i.e., $\Ca=0.35$, $0.22$ and $0.24$ for disc, square, and rectangle, respectively.
All cases take on strongly stretched conformations, and in addition, the rectangle takes on an asymmetric shape: reflection symmetry across the extensional axis is broken. The following paragraphs describe the parameter dependence in further detail.

\begin{figure*}[!htb]
    \centering
    \captionsetup{justification=raggedright}
    \includegraphics[width=\textwidth]{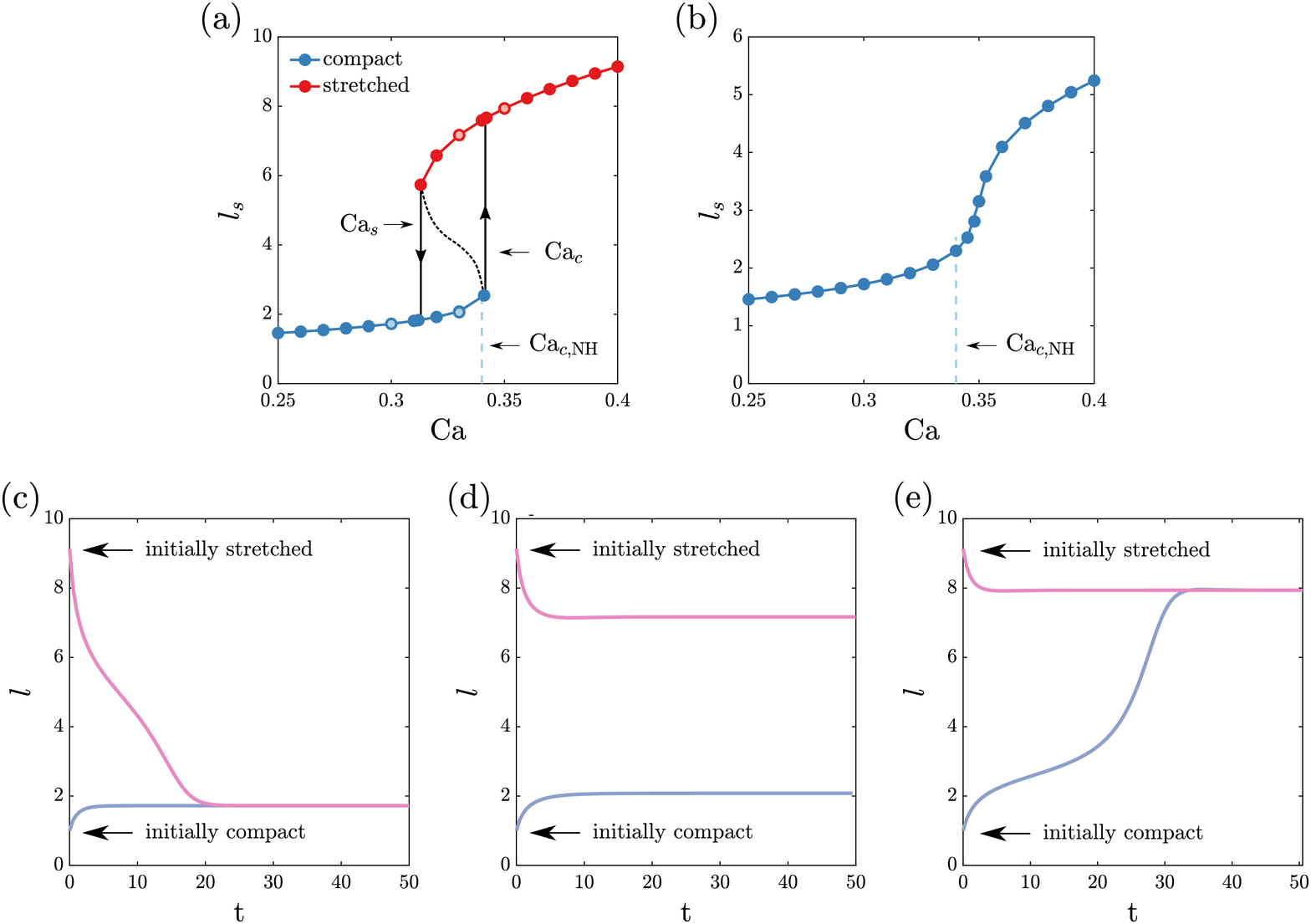}
    \caption{(a) Final stretched length $l_s$ vs. $\Ca$ for a disc with $c = 2\times 10^{-5}$ in planar extensional flow. Blue symbols and curves represent a compact final conformation and  red represent a stretched final conformation. The  points filled with light color correspond to the transient evolution shown in Figure \ref{fig:resp_disc_csh}c-e. The black dashed line indicates the unstable steady state. The light blue dashed line shows the critical $\Ca$ with neo-Hookean elasticity. $\Ca_c$ and $\Ca_s$ are two critical $\Ca$ marked besides the arrows. (b) Final stretched length $l_s$ vs. $\Ca$ for a disc with $c = 1\times 10^{-4}$ in planar extensional flow. (c)-(e): Examples of stretched length evolution of a disc from either a stretched state or a compact state with $c = 2\times 10^{-5}$ in planar extensional flow: (c) $\Ca = 0.3$ (d) $\Ca = 0.33$ (e) $\Ca = 0.35$.}
    \label{fig:resp_disc_csh}
\end{figure*}

Figure \ref{fig:resp_disc_csh}a shows steady state stretching length of a disc vs.~$\Ca$ for $c = 2\times 10^{-5}$. At sufficiently low $\Ca$  the only steady state is compact (blue). Once $\Ca$ exceeds a critical value $\Ca_c$, the compact steady state branch loses existence and the length evolves to a stretched state (red), as indicated by the upward black arrow. This type of transition is known as a saddle-node bifurcation. The critical value $\Ca_c$ splits the result into two branches in the parameter space: a compact branch and the stretched branch. Meanwhile, the discontinuity in the stretched length suggests the existence of a bistable region that corresponds to hysteresis as $\Ca$ is increased or decreased quasistatically.  Indeed this is the case -- if we use an initial condition corresponding to a stretched state, then as long as $\Ca$ is greater than some lower limit, denoted $\Ca_s$, then the final state will be highly stretched. For $\Ca<\Ca_s$, the stretched steady state branch loses existence, and an initially stretched sheet will relax to a compact steady state, as indicated by the downward black arrow. In general, this type of bistable behavior implies the existence of a branch of unstable steady states intermediate between the upper and lower ones. This branch is indicated schematically by the black dashed lines -- we have not explicitly computed it.  Figures \ref{fig:resp_disc_csh}c-e show the evolution of either an initially compact or an initially stretched state for $\Ca<\Ca_s$, $\Ca_s<\Ca<\Ca_c$, and $\Ca>\Ca_c$, respectively. 

In polymer dynamics, the origin of coil-stretch hysteresis is understood to be the conformation-dependent hydrodynamic interactions. The situation here is somewhat analogous. 
When the sheet remains at a compact state, the strain is relatively small and there is not a substantial change in surface area for the fluid to exert stresses on.  
In the stretched state, however, the surface area is substantially larger, leading to larger forces on the sheet from the flow, preventing relaxation back to a compact state. Consequently, the sheet maintains a deformed state that is larger in size than the compact state under the same $\Ca$. In Section \ref{sec:Brinkman}, we verify this picture by examination of an artificial model in which hydrodynamic interactions are artificially screened out.
  
\begin{figure*}[!htb]
    \centering
    \captionsetup{justification=raggedright}
    \includegraphics[width=\textwidth]{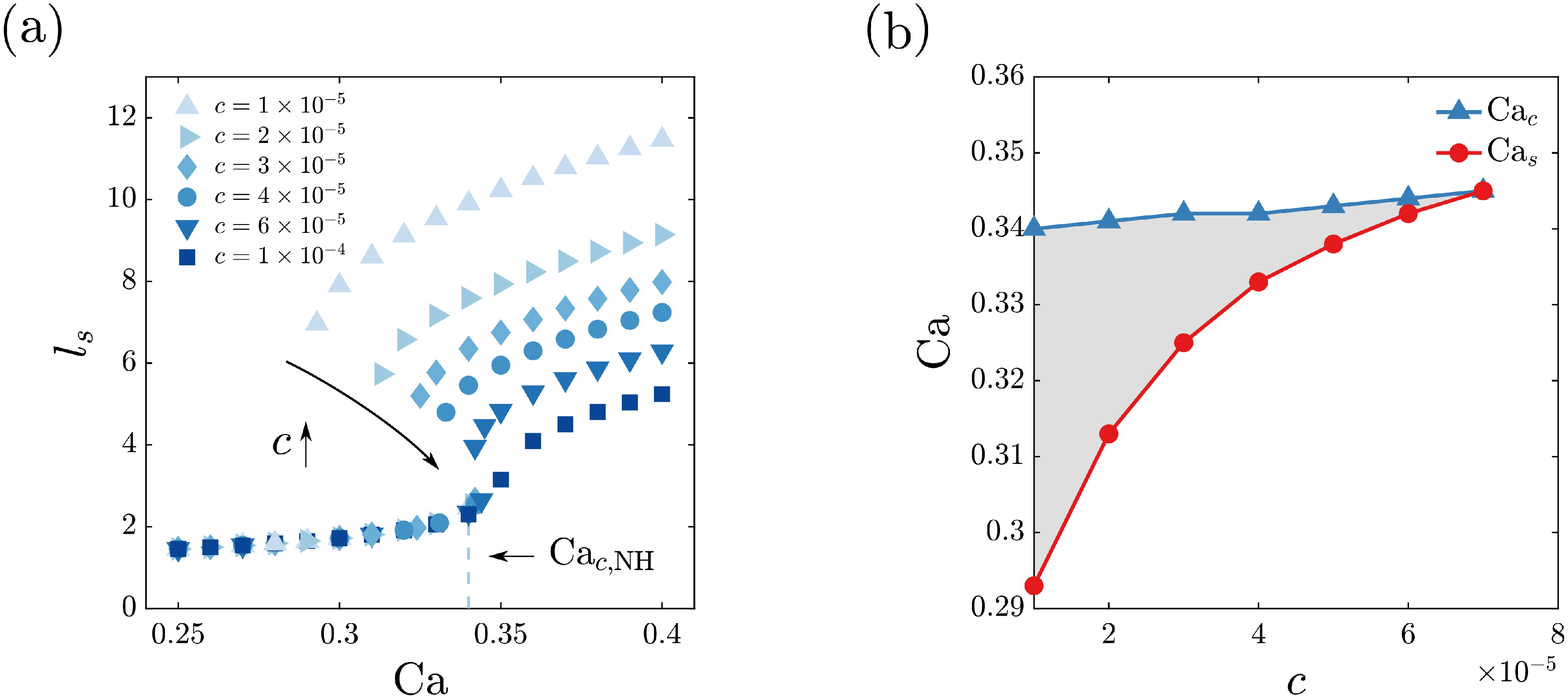}
    \caption{(a) Final stretched length $l_s$ vs. $\Ca$ for a disc in planar extensional flow with various $c$. All $c$ have a similar (overlapped) compact branch. The blue dashed line refers to $\Ca_s$ with neo-Hookean elasticity. (b) Phase diagram of $\Ca_c$ and $\Ca_s$ vs.~$c$ in planar extensional flow. The gray area marks the parameter regime where hysteresis can be observed.}
    \label{fig:resp_disc_all}
\end{figure*}

This bistability phenomenon only exists for small $c$. The steady-state stretch for $c = 1\times 10^{-4}$ is shown in Figure \ref{fig:resp_disc_csh}b; while there is a sharp increase in $l_s$ as $\Ca$ increases, there no longer exists a discontinuous transition or a multiplicity region. 
Figure \ref{fig:resp_disc_all}a shows the steady states over a range of $c$. All choices of $c$ share a similar compact branch, as the cubic energy term does not play a role when the strain is small. The discontinuity between the compact and the stretched branch becomes more evident with decreasing $c$, where hysteresis arises.  
\MDGrevise{In the linear polymer case,  bistable behavior in extension is only observed for very long chain -- highly extensible -- polymers, where the equilibrium coil size and the contour length differ by orders of magnitude. In the present case, the extensibility is represented by the nonlinearity parameter $c$ in the strain energy. Thinking of the sheet as a crosslinked polymer network, small $c$ roughly corresponds to a material with a large number of polymer segments between crosslinks: $c=0$ is the neo-Hookean limit where the material can stretch without limit. As $c$ increases, large deformations are increasingly penalized, as in the case of a linear chain with finite extensibility. Consistent with the linear polymer case, for the sheets, bistability vanishes as extensibility decreases ($c$ increases). }

Figure \ref{fig:resp_disc_all}b summarizes the values of $\Ca_c$ and $\Ca_s$ as a function of $c$. Hysteresis exists in the gray region between the two curves. Note that $\Ca_c$ increases only slightly with $c$, as the nonlinear elastic behavior only has a small influence for a compact state. The $\Ca_s$ branch increases faster with $c$ as a result of the increasing energy penalty for a stretched state as $c$ increases. Based on the chosen mesh resolution, we note that $\Ca_c$ and $\Ca_s$ are not sensitive to further increase in resolution; increasing the number of elements by 10$\%$ shifts $\Ca_s$ and $\Ca_c$ in the third decimal place.

\begin{figure*}[!htb]
    \centering
    \captionsetup{justification=raggedright}
    \includegraphics[width=\textwidth]{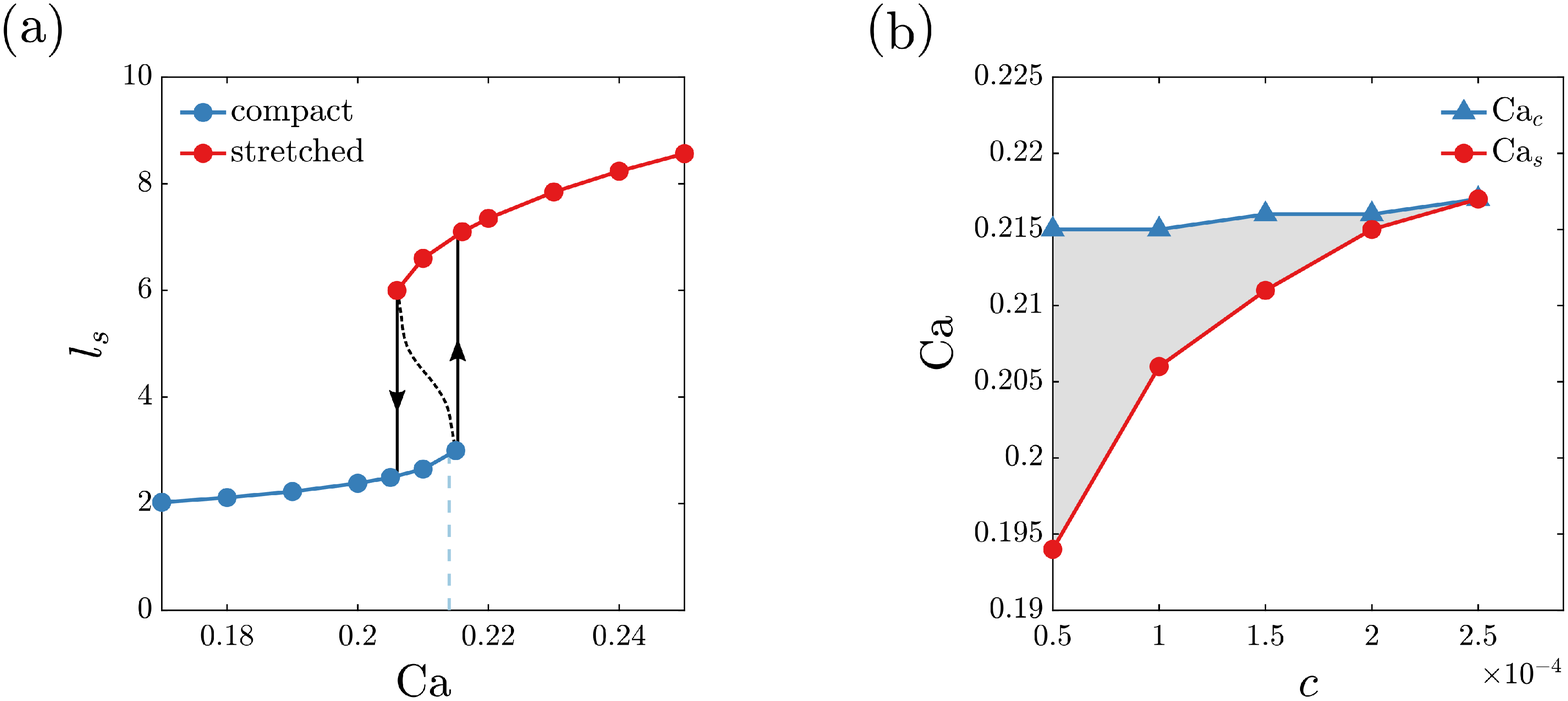}
    \caption{(a) Bifurcation diagram of stretched length $l_s$ vs. $\Ca$ for a square in planar extensional flow, $c = 1\times 10^{-4}$. (b) Phase diagram of  $\Ca_c$ and $\Ca_s$ vs $c$ for a square in planar extensional flow, with the gray area indicating the hysteresis region.}
    \label{fig:resp_sqr_csh}
\end{figure*}

Since we found compact-stretched transition and hysteresis behavior for a disc sheet in planar extensional flow, we expect a similar transition to be found for a different sheet geometries. We now briefly present  the behavior of a square in planar extensional flow. As previously mentioned, the square shape preferentially orients with diagonally opposed corners in the extension direction (Figure \ref{fig:resp_steady_state_compact}b and Figure \ref{fig:resp_steady_state_stretched}b) for both the compact and the stretched state. We provide an example of compact-stretched transition and hysteresis behavior for a square in Figure \ref{fig:resp_sqr_csh}a, for $c=1\times 10^{-4}$. 
Figure \ref{fig:resp_sqr_csh}b shows the region in $\Ca-c$ space where hysteresis can be found.

\begin{figure*}[!htb]
    \centering
    \captionsetup{justification=raggedright}
    \includegraphics[width=\textwidth]{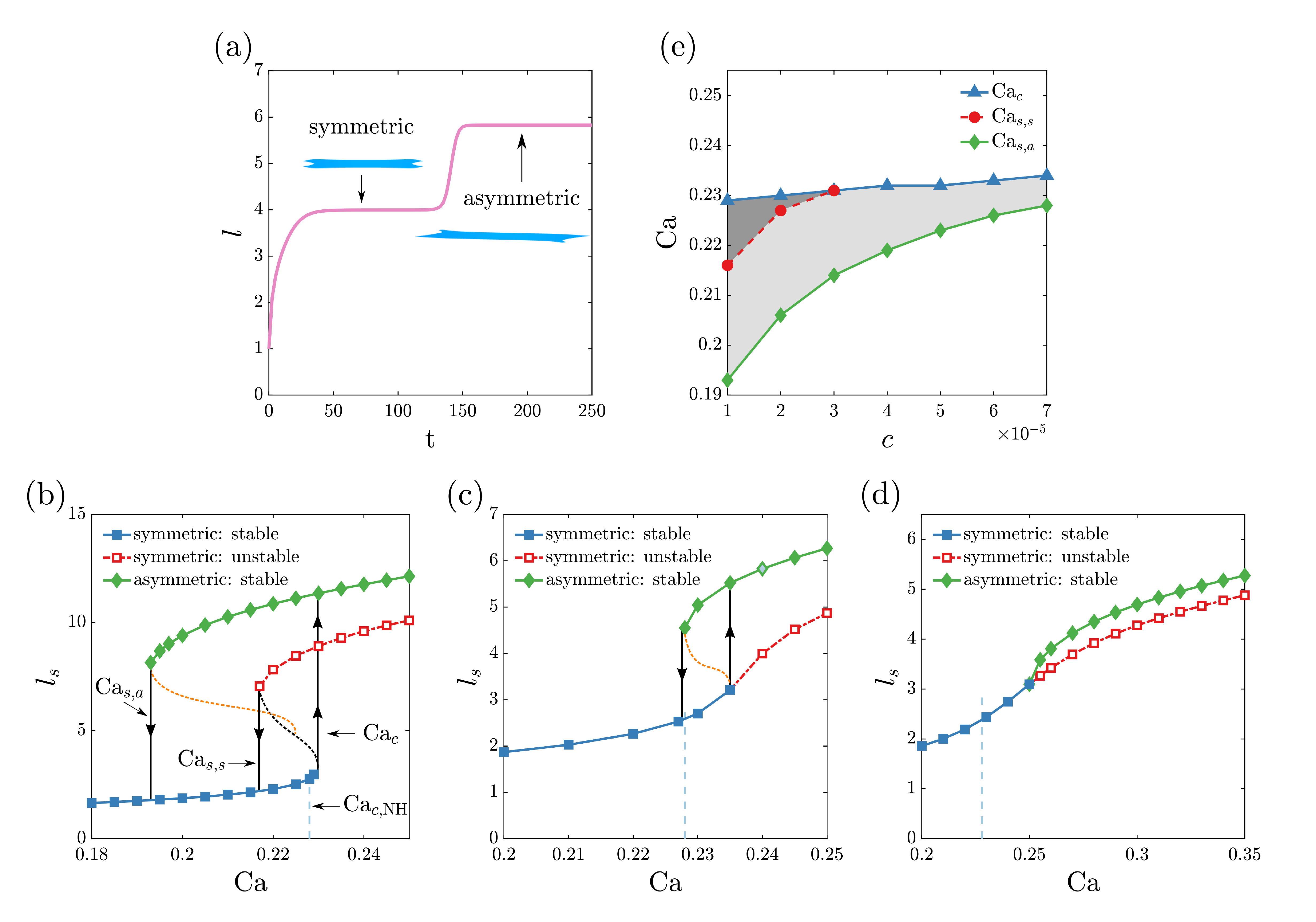}
    \caption{(a) Transient stretched length evolution of a rectangle in planar extensional flow, $c = 7\times 10^{-5}$, $\Ca = 0.24$. (b) Bifurcation diagram of $l_s$ vs. $\Ca$ for a rectangle with $c = 1\times 10^{-5}$.(c) Bifurcation diagram of $l_s$ vs. $\Ca$ for a rectangle sheet with $c = 7\times 10^{-5}$. The data point filled with light color corresponds to the evolution shown in Figure \ref{fig:resp_rect_csh}a (d) Bifurcation diagram of $l_s$ vs. $\Ca$ for a rectangle with $c = 3\times 10^{-4}$. (e) Phase diagram $\Ca_c$ and $\Ca_s$ vs $c$ for a rectangle in planar extensional flow; here $\Ca_{s,a}$ represents the stretched critical $\Ca$ for an asymmetric conformation, and $\Ca_{s,s}$ represents the stretched critical $\Ca$ for a symmetric conformation. The dark gray area indicates the hysteresis region  for symmetric states and the light gray area marks the hysteresis region without symmetry constraints, so the stretched state is tilted.}
    \label{fig:resp_rect_csh}
\end{figure*}

In contrast, the rectangle is less symmetric than the disc or the square. From the previous steady-state analysis, we have already noted the existence of an asymmetric stretched steady state. This result indicates that there exists a symmetry breaking bifurcation that interacts with the bifurcation behavior shown above. Figure \ref{fig:resp_rect_csh}a illustrates the length evolution for a rectangle initially in its rest state and oriented with the extension direction. This sheet is initially stretched by the flow symmetrically relative to the $x-z$ plane until it reaches a symmetric stretched state. The sheet stays at this unstable symmetric state for some time, but then evolves to a tilted asymmetric stable conformation. This observation suggests that there is a stretched symmetric state that is unstable with respect to symmetry-breaking perturbations, leading the final steady state to be asymmetric.

Figures \ref{fig:resp_rect_csh}b and c shows bifurcation diagrams for $c=1\times 10^{-5}$ and $7\times 10^{-5}$, respectively. (We can compute symmetric steady states that unstable in the full space by enforcing reflection symmetry.) In Figure \ref{fig:resp_rect_csh}b, multiplicity behavior exists for both the symmetric and the (unstable) symmetric stretched steady states. The bifurcation diagram indicates two saddle node bifurcations on the symmetric branch and a hypothesized subcritical symmetry-breaking (pitchfork) bifurcation that also undergoes a saddle-node bifurcation, yielding the stable stretched asymmetric state. This is the simplest bifurcation scenario consistent with the computational results -- the orange dashed curve has not been computed but is consistent with this scenario. For the larger value of $c$ considered in \ref{fig:resp_rect_csh}c, there is no bistability of symmetric solutions, but the asymmetric branch still bifurcates subcritically and turns around to yield a stable asymmetric stretched state. 
If we further increase $c$, bistability will vanish for the asymmetric solution and we obtain a supercritical bifurcation. We present a case in Figure \ref{fig:resp_rect_csh}d to show the existence of supercritical bifurcation.

Figure \ref{fig:resp_rect_csh}e summarizes the behavior, showing the parameter space where hysteresis arises for both the asymmetric and the symmetric stretched states. The conformational hysteresis disappears faster as $c$ increases for symmetric conformations than tilted ones.

\subsection{Biaxial extensional flow} 

\begin{figure}[!htb]
    \centering
    \captionsetup{justification=raggedright}
    \includegraphics[width=0.5\textwidth]{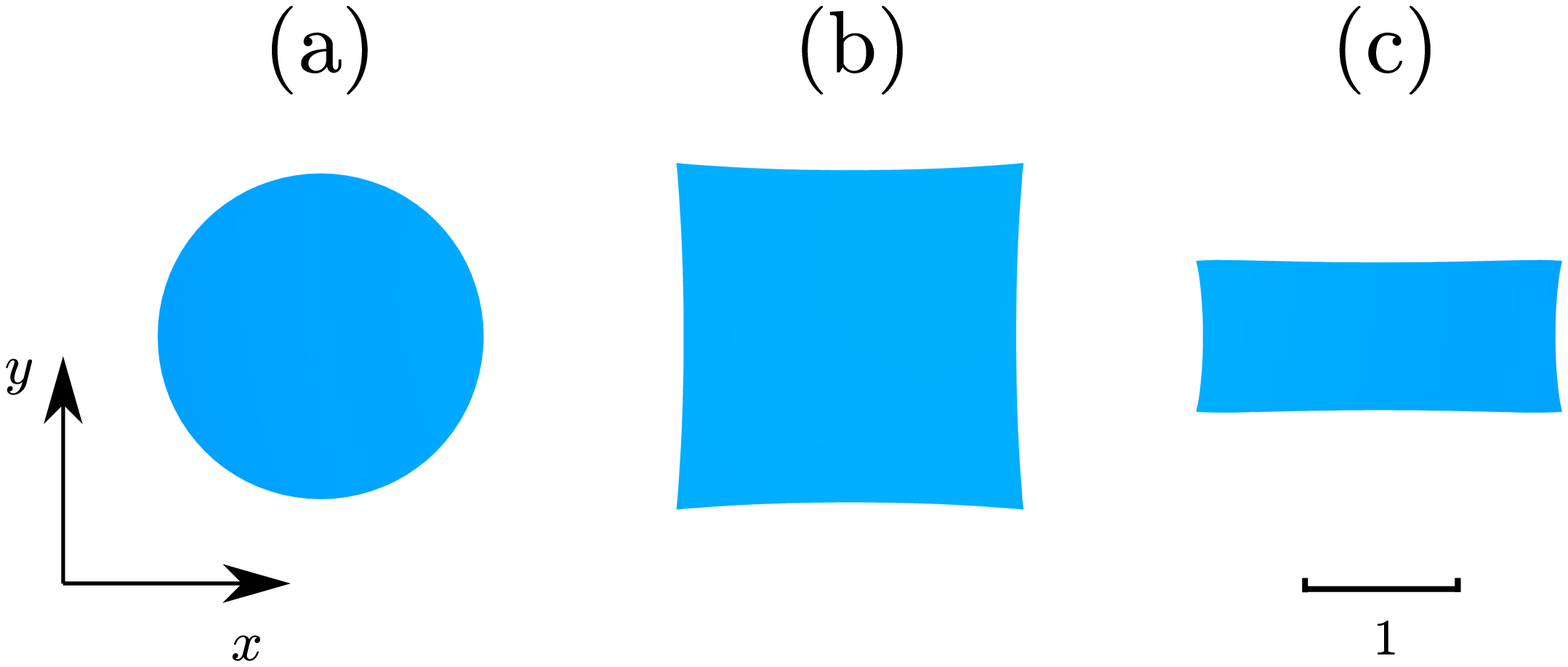}
    \caption{Steady-state conformation of a stable steady-state sheet with neo-Hookean elasticity in biaxial extensional flow at $\Ca = 0.2$: (a) disc. (b) square. (c) rectangle. Scale bar is 1 dimensionless length unit.}
    \label{fig:resb_steady_state_compact}
\end{figure}

This section briefly turns to the case of biaxial extension, in which we shall see many of the features described above in planar extension.  To begin with, similar to planar extension, the strong compressive flow inward along the $z$ direction suppresses any out-of-plane deformations, so at long times the  deformations only occur in the $x-y$ plane. Different from the planar extension case, due to the radial symmetry of the biaxial flow in the $x-y$ plane, there is no issue of a preferred orientation for a square and a rectangle. Examples of stable steady states are illustrated in Figure \ref{fig:resb_steady_state_compact}.

\begin{figure}[!htb]
    \centering
    \captionsetup{justification=raggedright}
    \includegraphics[width=0.42\textwidth]{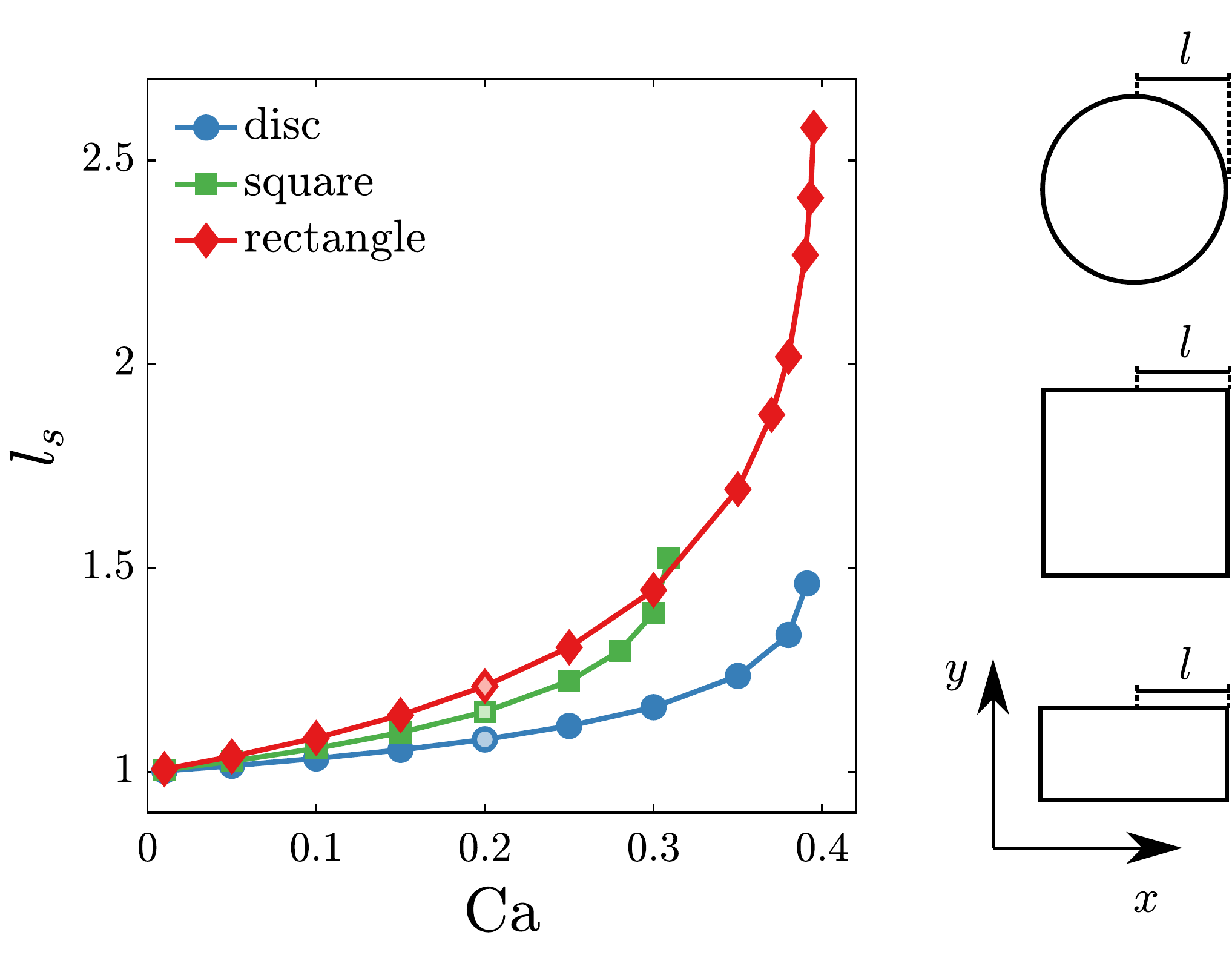}
    \caption{Steady-state stretched length $l_s$ vs. $\Ca$ for deformable sheets with neo-Hookean elasticity in biaxial extensional flow. The data points filled with light color correspond to the conformations shown in Figure \ref{fig:resb_steady_state_compact}. The critical capillary numbers are: disc: $\Ca_c = 0.391$, square: $\Ca_c = 0.309$, rectangle: $\Ca_c = 0.395$. The subplot indicates the length measured in each geometry.}
    \label{fig:resb_singular}
\end{figure}

\begin{figure*}[!htb]
    \centering
    \captionsetup{justification=raggedright}
    \includegraphics[width=\textwidth]{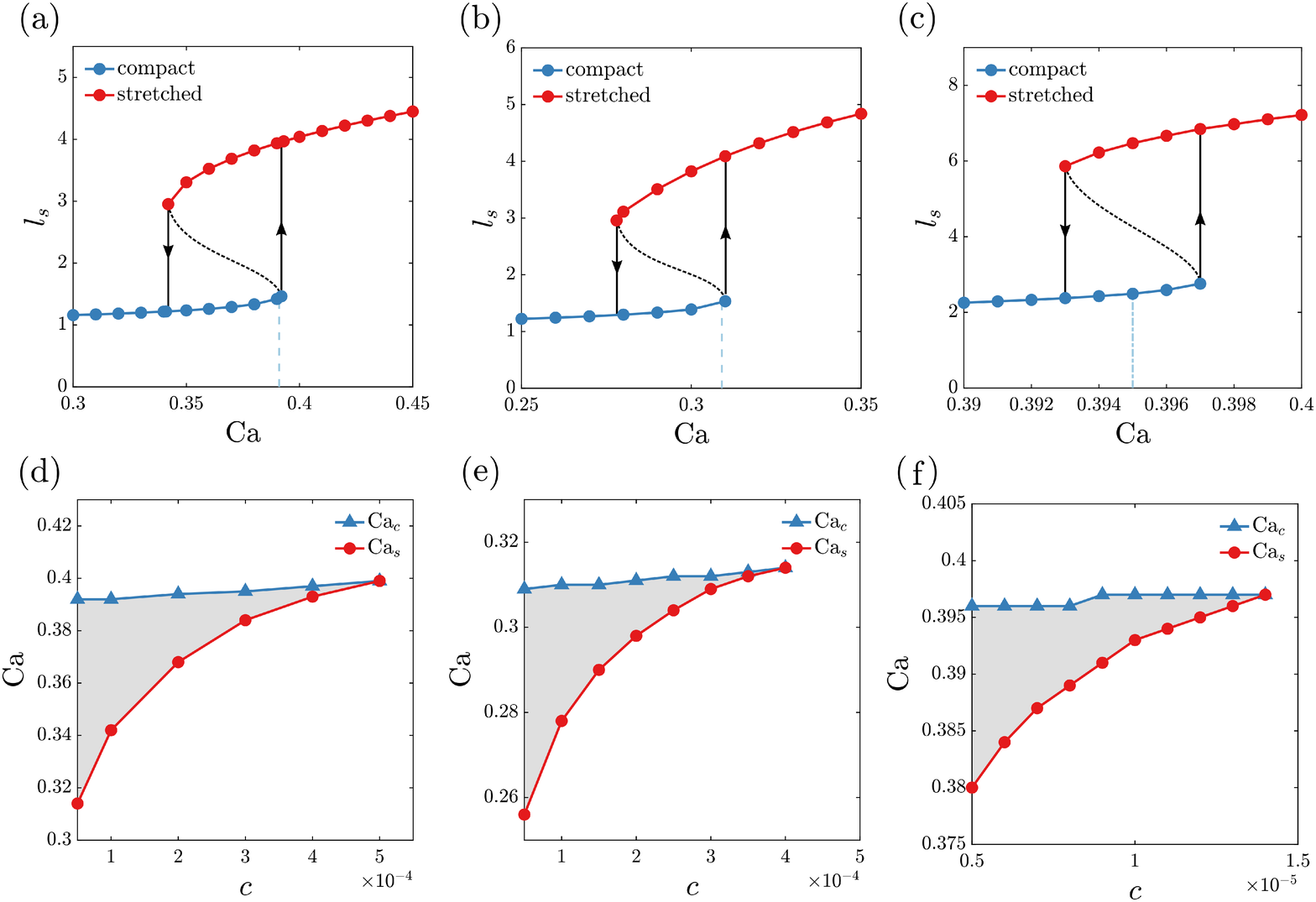}
    \caption{Bifurcation diagrams of $l_s$ vs. $\Ca$ for sheets of different shapes in biaxial extensional flow: (a) disc: $c = 1\times 10^{-4}$. (b) square: $c = 1\times 10^{-4}$, (c) rectangle: $c = 1\times 10^{-5}$.  The light blue dashed line refers to $\Ca_c$ with neo-Hookean elasticity. Phase diagram of the hysteresis regime in biaxial extensional flow: (d) disc, (e) square, (f) rectangle.}
    \label{fig:resb_all_csh}
\end{figure*}

As in planar extension, if neo-Hookean elasticity is used, there is again a singularity in stretching as $\Ca$ increases, as shown in Figure \ref{fig:resb_singular}. In general, $\Ca_c$ is larger compared to planar extension. With the Yeoh model, bistability behavior similar to that found above is again observed. In biaxial extension, all three shapes have a stretched state similar to its compact state, just more stretched. No symmetry-breaking has been observed. Figure \ref{fig:resb_all_csh}a-c show bifurcation diagrams for the disc, square and rectangle cases, respectively.

Figure \ref{fig:resb_all_csh}d-f give the corresponding phase diagrams, showing where where hysteresis occurs. Compared with planar extensional flow results, the disc and rectangle in biaxial flow have larger effective hysteresis regimes. Interestingly, the rectangle has a substantially smaller hysteresis regime in $c$. This may arise from the smaller extent of the rectangle in one of the two characteristic stretching directions of biaxial extension.  

\subsection{\MDGrevise{Effect of artificial hydrodynamic screening}}\label{sec:Brinkman}

\add{Recall that the presence of hysteresis in the coil-stretch transition  is due to the different hydrodynamic drag forces by the polymer in its coiled and stretched states\cite{de1974,schroeder2003}. To assess the role of hydrodynamic forces on bistability in the present case, we now consider an artificial model that weakens the hydrodynamic interaction between different parts of the sheet. In this model, we take the medium in which the sheet is moving to be a Brinkman porous medium rather than a simple viscous fluid. In a porous medium, hydrodynamic interactions decay quickly with distance, because the obstacles in the medium absorb momentum \cite{Graham2018}. With this model, the velocity field generated by point force $\mathbf{F}$ is determined not by the Stokeslet, but by the so-called Brinkmanlet:}

\begin{equation}
\begin{split}
\mathbf{B}(\mathbf{x})&=\frac{2}{8 \pi \eta \alpha^{2} r^{3}}\bigg[2\left(1-(1+\alpha r) e^{-\alpha r}\right) \frac{\mathbf{x} \mathbf{x}}{r^{2}}\\
&+\left(\left(1+\alpha r+\alpha^{2} r^{2}\right) e^{-\alpha r}-1\right)\left(\mathbf{I}-\frac{\mathbf{x} \mathbf{x}}{r^{2}}\right)\bigg]. 
\end{split}
\end{equation}

\add{Here a new length scale $\alpha^{-1}$, the Brinkman screening length, arises.  When $\alpha r \ll 1$, the Brinkmanlet reduces to the Stokeslet. When $\alpha r \gg 1$, the velocity decays as $r^{-3}$, as in a Darcy's law porous medium \cite{Graham2018}. In our model, we apply the Brinkmanlet, rather than the regularized Stokeslet, to the interactions between different nodes, and examine the effect of $\alpha$. 
In dimensionless form, $\alpha$ is scaled with sheet size $a$, so once $\alpha>1$, hydrodynamic will be important only between nearby positions on the sheet.} 

\begin{figure*}[!htb]
    \centering
    \captionsetup{justification=raggedright}
    \includegraphics[width=0.8\textwidth]{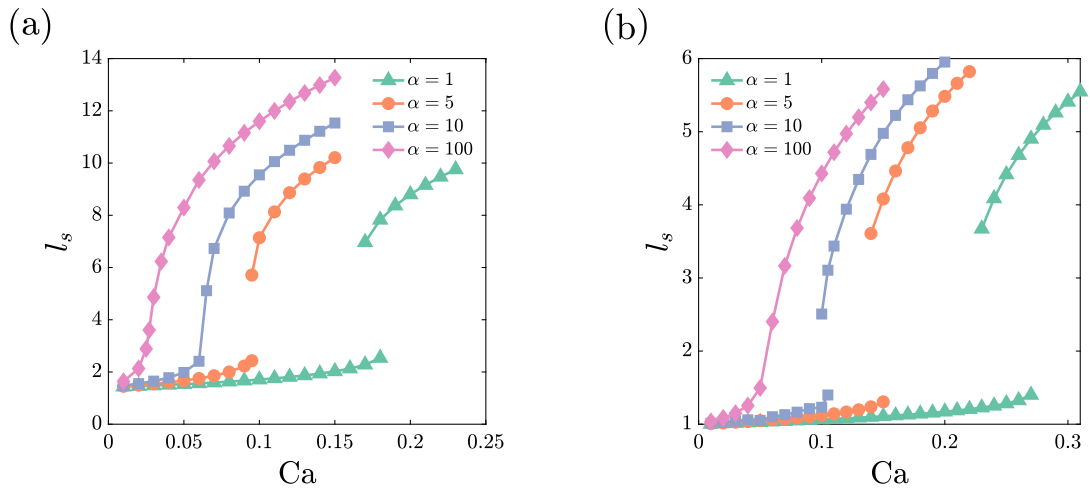}
    \caption{\add{Bifurcation diagrams of $l_s$ vs. $\Ca$ for a square sheet ($c = 1\times 10^{-4}$) with increasing $\alpha$: (a) planar extensional flow. (b) biaxial extensional flow.}}
    \label{fig:res_alpha}
\end{figure*}

\add{We take a square ($c = 1\times 10^{-4}$) in both planar and biaxial extension as an example to illustrate how the HI influences the hysteresis behavior. Figure \ref{fig:res_alpha} shows the steady state stretching length vs. $\Ca$ for multiple choices of $\alpha$. Compared with the case $\alpha = 0$ (Figure \ref{fig:resp_sqr_csh}a for planar extension and Figure \ref{fig:resb_all_csh}b for biaxial extension at $c = 1 \times 10^{-4}$), the bistable region becomes narrower and shifts towards smaller $\Ca$ as $\alpha$ increases. When $\alpha = 100$, the hysteresis region vanishes and we obtain a smooth monostable curve. We conclude that, as with linear polymers, bistable behaviors arise from the hydrodynamic interactions between different parts of the sheet surface.}

\section{Conclusions}\label{sec:conclusion}
In this study, we systematically explored the dynamics of freely suspended elastic sheets with three different rest shapes in planar and biaxial extensional flows. In both flow fields, the sheet always takes on a flat steady conformation in the stretching plane. If neo-Hookean elasticity is applied, we observed a weakly stretched ``compact" sheet conformation at low $\Ca$, while above a critical $\Ca_c$ the degree of stretching diverges. With a nonlinear Yeoh elasticity model, this singularity vanishes, but for sufficiently small nonlinearity parameter $c$, we observe a ``compact-stretched" transition, where the sheet suddenly becomes highly stretched once a critical capillary number is exceeded. 
The discontinuity in stretched length marks a bistable or hysteretic regime, where the sheet has either a compact or a stretched conformation based on its deformation history. This behavior occurs for all three rest shapes considered. \add{The origin of such bistable behavior is a result of hydrodynamic interaction between different parts of the sheet. The dynamics become monostable if  hydrodynamic interactions are artificially screened out.}
The highly nonlinear behavior predicted here may play a substantial role in the behavior and performance of flow processes involving soft sheet-like particles.



\section*{Conflicts of interest}
There are no conflicts to declare.

\section*{Acknowledgements}
This material is based on work supported by the National Science Foundation under grant No.~CBET-1604767. This work used the Extreme Science and Engineering Discovery Environment (XSEDE) \cite{towns2014xsede} SDSC Dell Cluster with Intel Haswell Processors (Comet) through allocations TG-CTS190001 and TG-MCB190100.



\balance


\bibliography{rsc,cite} 

\providecommand*{\mcitethebibliography}{\thebibliography}
\csname @ifundefined\endcsname{endmcitethebibliography}
{\let\endmcitethebibliography\endthebibliography}{}
\begin{mcitethebibliography}{39}
\providecommand*{\natexlab}[1]{#1}
\providecommand*{\mciteSetBstSublistMode}[1]{}
\providecommand*{\mciteSetBstMaxWidthForm}[2]{}
\providecommand*{\mciteBstWouldAddEndPuncttrue}
  {\def\EndOfBibitem{\unskip.}}
\providecommand*{\mciteBstWouldAddEndPunctfalse}
  {\let\EndOfBibitem\relax}
\providecommand*{\mciteSetBstMidEndSepPunct}[3]{}
\providecommand*{\mciteSetBstSublistLabelBeginEnd}[3]{}
\providecommand*{\EndOfBibitem}{}
\mciteSetBstSublistMode{f}
\mciteSetBstMaxWidthForm{subitem}
{(\emph{\alph{mcitesubitemcount}})}
\mciteSetBstSublistLabelBeginEnd{\mcitemaxwidthsubitemform\space}
{\relax}{\relax}

\bibitem[Lee \emph{et~al.}(2013)Lee, Pham, Lawrence, Lee, Parkos, Emrick, and
  Crosby]{lee2013}
D.~Y. Lee, J.~T. Pham, J.~Lawrence, C.~H. Lee, C.~Parkos, T.~Emrick and A.~J.
  Crosby, \emph{Advanced materials}, 2013, \textbf{25}, 1248--1253\relax
\mciteBstWouldAddEndPuncttrue
\mciteSetBstMidEndSepPunct{\mcitedefaultmidpunct}
{\mcitedefaultendpunct}{\mcitedefaultseppunct}\relax
\EndOfBibitem
\bibitem[Ni \emph{et~al.}(2015)Ni, Huang, Chen, Chen, Hsu, Li, Pochan, Zhang,
  Cheng, and Dong]{ni2015}
B.~Ni, M.~Huang, Z.~Chen, Y.~Chen, C.-H. Hsu, Y.~Li, D.~Pochan, W.-B. Zhang,
  S.~Z. Cheng and X.-H. Dong, \emph{Journal of the American Chemical Society},
  2015, \textbf{137}, 1392--1395\relax
\mciteBstWouldAddEndPuncttrue
\mciteSetBstMidEndSepPunct{\mcitedefaultmidpunct}
{\mcitedefaultendpunct}{\mcitedefaultseppunct}\relax
\EndOfBibitem
\bibitem[So and Hayward(2017)]{so2017}
S.~So and R.~C. Hayward, \emph{ACS Applied Materials \& Interfaces}, 2017,
  \textbf{9}, 15785--15790\relax
\mciteBstWouldAddEndPuncttrue
\mciteSetBstMidEndSepPunct{\mcitedefaultmidpunct}
{\mcitedefaultendpunct}{\mcitedefaultseppunct}\relax
\EndOfBibitem
\bibitem[Jamal \emph{et~al.}(2011)Jamal, Zarafshar, and Gracias]{jamal2011}
M.~Jamal, A.~M. Zarafshar and D.~H. Gracias, \emph{Nature communications},
  2011, \textbf{2}, 1--6\relax
\mciteBstWouldAddEndPuncttrue
\mciteSetBstMidEndSepPunct{\mcitedefaultmidpunct}
{\mcitedefaultendpunct}{\mcitedefaultseppunct}\relax
\EndOfBibitem
\bibitem[Hubbard \emph{et~al.}(2017)Hubbard, Mailen, Zikry, Dickey, and
  Genzer]{hubbard2017}
A.~M. Hubbard, R.~W. Mailen, M.~A. Zikry, M.~D. Dickey and J.~Genzer,
  \emph{Soft Matter}, 2017, \textbf{13}, 2299--2308\relax
\mciteBstWouldAddEndPuncttrue
\mciteSetBstMidEndSepPunct{\mcitedefaultmidpunct}
{\mcitedefaultendpunct}{\mcitedefaultseppunct}\relax
\EndOfBibitem
\bibitem[Na \emph{et~al.}(2015)Na, Evans, Bae, Chiappelli, Santangelo, Lang,
  Hull, and Hayward]{na2015}
J.-H. Na, A.~A. Evans, J.~Bae, M.~C. Chiappelli, C.~D. Santangelo, R.~J. Lang,
  T.~C. Hull and R.~C. Hayward, \emph{Advanced Materials}, 2015, \textbf{27},
  79--85\relax
\mciteBstWouldAddEndPuncttrue
\mciteSetBstMidEndSepPunct{\mcitedefaultmidpunct}
{\mcitedefaultendpunct}{\mcitedefaultseppunct}\relax
\EndOfBibitem
\bibitem[Feinberg \emph{et~al.}(2007)Feinberg, Feigel, Shevkoplyas, Sheehy,
  Whitesides, and Parker]{feinberg2007}
A.~W. Feinberg, A.~Feigel, S.~S. Shevkoplyas, S.~Sheehy, G.~M. Whitesides and
  K.~K. Parker, \emph{Science}, 2007, \textbf{317}, 1366--1370\relax
\mciteBstWouldAddEndPuncttrue
\mciteSetBstMidEndSepPunct{\mcitedefaultmidpunct}
{\mcitedefaultendpunct}{\mcitedefaultseppunct}\relax
\EndOfBibitem
\bibitem[Chun \emph{et~al.}(2013)Chun, Kim, Shin, Kim, Spinks, Aliev, Baughman,
  and Kim]{chun2013free}
K.-Y. Chun, S.~H. Kim, M.~K. Shin, Y.~T. Kim, G.~M. Spinks, A.~E. Aliev, R.~H.
  Baughman and S.~J. Kim, \emph{Nanotechnology}, 2013, \textbf{24},
  165401\relax
\mciteBstWouldAddEndPuncttrue
\mciteSetBstMidEndSepPunct{\mcitedefaultmidpunct}
{\mcitedefaultendpunct}{\mcitedefaultseppunct}\relax
\EndOfBibitem
\bibitem[Zhao \emph{et~al.}(2019)Zhao, Guan, and Wu]{zhao2019highly}
Y.~Zhao, J.~Guan and S.~J. Wu, \emph{Macromolecular Rapid Communications},
  2019, \textbf{40}, 1900389\relax
\mciteBstWouldAddEndPuncttrue
\mciteSetBstMidEndSepPunct{\mcitedefaultmidpunct}
{\mcitedefaultendpunct}{\mcitedefaultseppunct}\relax
\EndOfBibitem
\bibitem[Gaharwar \emph{et~al.}(2011)Gaharwar, Dammu, Canter, Wu, and
  Schmidt]{gaharwar2011highly}
A.~K. Gaharwar, S.~A. Dammu, J.~M. Canter, C.-J. Wu and G.~Schmidt,
  \emph{Biomacromolecules}, 2011, \textbf{12}, 1641--1650\relax
\mciteBstWouldAddEndPuncttrue
\mciteSetBstMidEndSepPunct{\mcitedefaultmidpunct}
{\mcitedefaultendpunct}{\mcitedefaultseppunct}\relax
\EndOfBibitem
\bibitem[Liu \emph{et~al.}(2012)Liu, Chen, He, Zhao, Yang, and
  Wang]{liu2012synthesis}
J.~Liu, C.~Chen, C.~He, J.~Zhao, X.~Yang and H.~Wang, \emph{ACS nano}, 2012,
  \textbf{6}, 8194--8202\relax
\mciteBstWouldAddEndPuncttrue
\mciteSetBstMidEndSepPunct{\mcitedefaultmidpunct}
{\mcitedefaultendpunct}{\mcitedefaultseppunct}\relax
\EndOfBibitem
\bibitem[Mehta \emph{et~al.}(2018)Mehta, Jin, Stanciulescu, and
  Grande-Allen]{mehta2018engineering}
S.~M. Mehta, T.~Jin, I.~Stanciulescu and K.~J. Grande-Allen, \emph{Acta
  Biomaterialia}, 2018, \textbf{75}, 52--62\relax
\mciteBstWouldAddEndPuncttrue
\mciteSetBstMidEndSepPunct{\mcitedefaultmidpunct}
{\mcitedefaultendpunct}{\mcitedefaultseppunct}\relax
\EndOfBibitem
\bibitem[De~Gennes(1974)]{de1974}
P.~De~Gennes, \emph{The Journal of Chemical Physics}, 1974, \textbf{60},
  5030--5042\relax
\mciteBstWouldAddEndPuncttrue
\mciteSetBstMidEndSepPunct{\mcitedefaultmidpunct}
{\mcitedefaultendpunct}{\mcitedefaultseppunct}\relax
\EndOfBibitem
\bibitem[Schroeder \emph{et~al.}(2003)Schroeder, Babcock, Shaqfeh, and
  Chu]{schroeder2003}
C.~M. Schroeder, H.~P. Babcock, E.~S. Shaqfeh and S.~Chu, \emph{Science}, 2003,
  \textbf{301}, 1515--1519\relax
\mciteBstWouldAddEndPuncttrue
\mciteSetBstMidEndSepPunct{\mcitedefaultmidpunct}
{\mcitedefaultendpunct}{\mcitedefaultseppunct}\relax
\EndOfBibitem
\bibitem[Fuller and Leal(1981)]{fuller1981}
G.~Fuller and L.~Leal, \emph{Journal of Non-Newtonian Fluid Mechanics}, 1981,
  \textbf{8}, 271--310\relax
\mciteBstWouldAddEndPuncttrue
\mciteSetBstMidEndSepPunct{\mcitedefaultmidpunct}
{\mcitedefaultendpunct}{\mcitedefaultseppunct}\relax
\EndOfBibitem
\bibitem[Nafar~Sefiddashti \emph{et~al.}(2018)Nafar~Sefiddashti, Edwards, and
  Khomami]{nafar2018}
M.~H. Nafar~Sefiddashti, B.~J. Edwards and B.~Khomami, \emph{The Journal of
  Chemical Physics}, 2018, \textbf{148}, 141103\relax
\mciteBstWouldAddEndPuncttrue
\mciteSetBstMidEndSepPunct{\mcitedefaultmidpunct}
{\mcitedefaultendpunct}{\mcitedefaultseppunct}\relax
\EndOfBibitem
\bibitem[Kantsler \emph{et~al.}(2008)Kantsler, Segre, and
  Steinberg]{kantsler2008}
V.~Kantsler, E.~Segre and V.~Steinberg, \emph{Physical review letters}, 2008,
  \textbf{101}, 048101\relax
\mciteBstWouldAddEndPuncttrue
\mciteSetBstMidEndSepPunct{\mcitedefaultmidpunct}
{\mcitedefaultendpunct}{\mcitedefaultseppunct}\relax
\EndOfBibitem
\bibitem[Narsimhan \emph{et~al.}(2015)Narsimhan, Spann, and
  Shaqfeh]{narsimhan2015}
V.~Narsimhan, A.~P. Spann and E.~S. Shaqfeh, \emph{Journal of Fluid Mechanics},
  2015, \textbf{777}, 1--26\relax
\mciteBstWouldAddEndPuncttrue
\mciteSetBstMidEndSepPunct{\mcitedefaultmidpunct}
{\mcitedefaultendpunct}{\mcitedefaultseppunct}\relax
\EndOfBibitem
\bibitem[Kumar \emph{et~al.}(2020)Kumar, Richter, and Schroeder]{kumar2020}
D.~Kumar, C.~M. Richter and C.~M. Schroeder, \emph{Soft Matter}, 2020,
  \textbf{16}, 337--347\relax
\mciteBstWouldAddEndPuncttrue
\mciteSetBstMidEndSepPunct{\mcitedefaultmidpunct}
{\mcitedefaultendpunct}{\mcitedefaultseppunct}\relax
\EndOfBibitem
\bibitem[Shelley and Zhang(2011)]{shelley2011flapping}
M.~J. Shelley and J.~Zhang, \emph{Annual Review of Fluid Mechanics}, 2011,
  \textbf{43}, 449--465\relax
\mciteBstWouldAddEndPuncttrue
\mciteSetBstMidEndSepPunct{\mcitedefaultmidpunct}
{\mcitedefaultendpunct}{\mcitedefaultseppunct}\relax
\EndOfBibitem
\bibitem[Alben and Shelley(2008)]{alben2008flapping}
S.~Alben and M.~J. Shelley, \emph{Physical review letters}, 2008, \textbf{100},
  074301\relax
\mciteBstWouldAddEndPuncttrue
\mciteSetBstMidEndSepPunct{\mcitedefaultmidpunct}
{\mcitedefaultendpunct}{\mcitedefaultseppunct}\relax
\EndOfBibitem
\bibitem[Xu and Green(2014)]{xu2014}
Y.~Xu and M.~J. Green, \emph{The Journal of chemical physics}, 2014,
  \textbf{141}, 024905\relax
\mciteBstWouldAddEndPuncttrue
\mciteSetBstMidEndSepPunct{\mcitedefaultmidpunct}
{\mcitedefaultendpunct}{\mcitedefaultseppunct}\relax
\EndOfBibitem
\bibitem[Xu and Green(2015)]{xu2015}
Y.~Xu and M.~J. Green, \emph{Journal of Polymer Science Part B: Polymer
  Physics}, 2015, \textbf{53}, 1247--1253\relax
\mciteBstWouldAddEndPuncttrue
\mciteSetBstMidEndSepPunct{\mcitedefaultmidpunct}
{\mcitedefaultendpunct}{\mcitedefaultseppunct}\relax
\EndOfBibitem
\bibitem[Gravelle \emph{et~al.}(2020)Gravelle, Kamal, and
  Botto]{gravelle2020liquid}
S.~Gravelle, C.~Kamal and L.~Botto, \emph{The Journal of Chemical Physics},
  2020, \textbf{152}, 104701\relax
\mciteBstWouldAddEndPuncttrue
\mciteSetBstMidEndSepPunct{\mcitedefaultmidpunct}
{\mcitedefaultendpunct}{\mcitedefaultseppunct}\relax
\EndOfBibitem
\bibitem[Kamal \emph{et~al.}(2020)Kamal, Gravelle, and
  Botto]{kamal2020hydrodynamic}
C.~Kamal, S.~Gravelle and L.~Botto, \emph{Nature communications}, 2020,
  \textbf{11}, 1--10\relax
\mciteBstWouldAddEndPuncttrue
\mciteSetBstMidEndSepPunct{\mcitedefaultmidpunct}
{\mcitedefaultendpunct}{\mcitedefaultseppunct}\relax
\EndOfBibitem
\bibitem[Dutta and Graham(2017)]{dutta2017}
S.~Dutta and M.~D. Graham, \emph{Soft Matter}, 2017, \textbf{13},
  2620--2633\relax
\mciteBstWouldAddEndPuncttrue
\mciteSetBstMidEndSepPunct{\mcitedefaultmidpunct}
{\mcitedefaultendpunct}{\mcitedefaultseppunct}\relax
\EndOfBibitem
\bibitem[Yeoh(1993)]{yeoh1993}
O.~H. Yeoh, \emph{Rubber Chemistry and technology}, 1993, \textbf{66},
  754--771\relax
\mciteBstWouldAddEndPuncttrue
\mciteSetBstMidEndSepPunct{\mcitedefaultmidpunct}
{\mcitedefaultendpunct}{\mcitedefaultseppunct}\relax
\EndOfBibitem
\bibitem[Fedosov \emph{et~al.}(2010)Fedosov, Caswell, and
  Karniadakis]{Fedosov2010}
D.~A. Fedosov, B.~Caswell and G.~E. Karniadakis, \emph{Computer Methods in
  Applied Mechanics and Engineering}, 2010, \textbf{199}, 1937--1948\relax
\mciteBstWouldAddEndPuncttrue
\mciteSetBstMidEndSepPunct{\mcitedefaultmidpunct}
{\mcitedefaultendpunct}{\mcitedefaultseppunct}\relax
\EndOfBibitem
\bibitem[Fedosov(2010)]{fedosov2010multiscale}
D.~A. Fedosov, \emph{Multiscale modeling of blood flow and soft matter},
  Citeseer, 2010\relax
\mciteBstWouldAddEndPuncttrue
\mciteSetBstMidEndSepPunct{\mcitedefaultmidpunct}
{\mcitedefaultendpunct}{\mcitedefaultseppunct}\relax
\EndOfBibitem
\bibitem[Timoshenko and Woinowsky-Krieger(1959)]{timoshenko1959theory}
S.~P. Timoshenko and S.~Woinowsky-Krieger, \emph{Theory of plates and shells},
  McGraw-hill, 1959\relax
\mciteBstWouldAddEndPuncttrue
\mciteSetBstMidEndSepPunct{\mcitedefaultmidpunct}
{\mcitedefaultendpunct}{\mcitedefaultseppunct}\relax
\EndOfBibitem
\bibitem[Liu \emph{et~al.}(2012)Liu, Metcalf, Robinson, Houston, and
  Scarpa]{liu2012shear}
X.~Liu, T.~H. Metcalf, J.~T. Robinson, B.~H. Houston and F.~Scarpa, \emph{Nano
  letters}, 2012, \textbf{12}, 1013--1017\relax
\mciteBstWouldAddEndPuncttrue
\mciteSetBstMidEndSepPunct{\mcitedefaultmidpunct}
{\mcitedefaultendpunct}{\mcitedefaultseppunct}\relax
\EndOfBibitem
\bibitem[Charrier \emph{et~al.}(1989)Charrier, Shrivastava, and
  Wu]{Charrier1989}
J.~Charrier, S.~Shrivastava and R.~Wu, \emph{The Journal of Strain Analysis for
  Engineering Design}, 1989, \textbf{24}, 55--74\relax
\mciteBstWouldAddEndPuncttrue
\mciteSetBstMidEndSepPunct{\mcitedefaultmidpunct}
{\mcitedefaultendpunct}{\mcitedefaultseppunct}\relax
\EndOfBibitem
\bibitem[Pappu and Bagchi(2008)]{Pappu:2008in}
V.~Pappu and P.~Bagchi, \emph{Comput Biol Med}, 2008, \textbf{38},
  738--753\relax
\mciteBstWouldAddEndPuncttrue
\mciteSetBstMidEndSepPunct{\mcitedefaultmidpunct}
{\mcitedefaultendpunct}{\mcitedefaultseppunct}\relax
\EndOfBibitem
\bibitem[Kumar and Graham(2012)]{Kumar:2012ev}
A.~Kumar and M.~D. Graham, \emph{Journal of Computational Physics}, 2012,
  \textbf{231}, 6682--6713\relax
\mciteBstWouldAddEndPuncttrue
\mciteSetBstMidEndSepPunct{\mcitedefaultmidpunct}
{\mcitedefaultendpunct}{\mcitedefaultseppunct}\relax
\EndOfBibitem
\bibitem[Cortez \emph{et~al.}(2005)Cortez, Fauci, and Medovikov]{Cortez2005}
R.~Cortez, L.~Fauci and A.~Medovikov, \emph{Physics of Fluids}, 2005,
  \textbf{17}, 031504\relax
\mciteBstWouldAddEndPuncttrue
\mciteSetBstMidEndSepPunct{\mcitedefaultmidpunct}
{\mcitedefaultendpunct}{\mcitedefaultseppunct}\relax
\EndOfBibitem
\bibitem[Hern{\'a}ndez-Ortiz \emph{et~al.}(2007)Hern{\'a}ndez-Ortiz, de~Pablo,
  and Graham]{hernandez2007}
J.~P. Hern{\'a}ndez-Ortiz, J.~J. de~Pablo and M.~D. Graham, \emph{Physical
  review letters}, 2007, \textbf{98}, 140602\relax
\mciteBstWouldAddEndPuncttrue
\mciteSetBstMidEndSepPunct{\mcitedefaultmidpunct}
{\mcitedefaultendpunct}{\mcitedefaultseppunct}\relax
\EndOfBibitem
\bibitem[Graham(2018)]{Graham2018}
M.~D. Graham, \emph{Microhydrodynamics, Brownian motion, and complex fluids},
  Cambridge University Press, 2018, vol.~58\relax
\mciteBstWouldAddEndPuncttrue
\mciteSetBstMidEndSepPunct{\mcitedefaultmidpunct}
{\mcitedefaultendpunct}{\mcitedefaultseppunct}\relax
\EndOfBibitem
\bibitem[Happel and Brenner(2012)]{happel2012}
J.~Happel and H.~Brenner, \emph{Low Reynolds number hydrodynamics: with special
  applications to particulate media}, Springer Science \& Business Media, 2012,
  vol.~1\relax
\mciteBstWouldAddEndPuncttrue
\mciteSetBstMidEndSepPunct{\mcitedefaultmidpunct}
{\mcitedefaultendpunct}{\mcitedefaultseppunct}\relax
\EndOfBibitem
\bibitem[Towns \emph{et~al.}(2014)Towns, Cockerill, Dahan, Foster, Gaither,
  Grimshaw, Hazlewood, Lathrop, Lifka, Peterson,\emph{et~al.}]{towns2014xsede}
J.~Towns, T.~Cockerill, M.~Dahan, I.~Foster, K.~Gaither, A.~Grimshaw,
  V.~Hazlewood, S.~Lathrop, D.~Lifka, G.~D. Peterson \emph{et~al.},
  \emph{Computing in science \& engineering}, 2014, \textbf{16}, 62--74\relax
\mciteBstWouldAddEndPuncttrue
\mciteSetBstMidEndSepPunct{\mcitedefaultmidpunct}
{\mcitedefaultendpunct}{\mcitedefaultseppunct}\relax
\EndOfBibitem
\end{mcitethebibliography}
\bibliographystyle{rsc} 

\end{document}